\documentclass[prb,twocolumn,floatfix,notitlepage,superscriptaddress,longtable,nofootinbib]{revtex4-2}
\usepackage{amsmath}
\usepackage{lipsum} 
\usepackage{verbatim}
\usepackage{epsfig}
\usepackage{subfigure}
\usepackage{graphicx}
\usepackage{amsfonts}
\usepackage[final]{showlabels}
\usepackage{enumitem}
\usepackage[figuresright]{rotating}
\usepackage{amssymb}
\usepackage{amsmath}
\usepackage{psfrag}
\usepackage{bm}
\usepackage[colorlinks,linkcolor=blue,anchorcolor=blue,citecolor=blue,urlcolor=blue]{hyperref}
\usepackage[version=4]{mhchem}
\usepackage[svgnames]{xcolor}

\def\be{\begin{equation}} \def\ee{\end{equation}}
\def\bea{\begin{eqnarray}} \def\eea{\end{eqnarray}}

\newcommand{\ket}[1]{| #1 \rangle}

\def\bpm{\begin{pmatrix}} \def\epm{\end{pmatrix}}

\DeclareMathOperator{\Tr}{Tr}

\definecolor{Qicolor}{RGB}{3, 136, 252}

\makeatletter
\newcommand*{\balancecolsandclearpage}{%
  \close@column@grid
  \clearpage
}
\makeatother

\begin{document}

\title{
Surprises in the Deep Hilbert Space of all-to-all systems:\\ From super-exponential scrambling to slow entanglement growth}


\author{Zihao Qi}
\affiliation{Department of Physics, Mathematics, and Astronomy, California Institute of Technology, Pasadena, CA 91125, USA.}
\affiliation{Department of Physics, Cornell University, Ithaca, NY 14853, USA.}

\author{Thomas Scaffidi}
\affiliation{Department of Physics, University of California, Irvine, Irvine, CA 92697, USA}
\affiliation{Department of Physics, University of Toronto, 60 St. George Street, Toronto, Ontario, M5S 1A7, Canada}

\author{Xiangyu Cao}
\affiliation{Laboratoire de Physique de l'\'Ecole normale sup\'erieure, ENS, Universit\'e PSL, CNRS, Sorbonne Universit\'e, Universit\'e Paris Cit\'e, F-75005 Paris, France}

\date{\today}

\begin{abstract}
The quantum dynamics of spin systems with uniform all-to-all interaction are often studied in the totally symmetric space (TSS) of maximal total spin. However the TSS states are atypical in the full many-body Hilbert space.   In this work, we explore several aspects of the all-to-all quantum dynamics away from the TSS, and reveal surprising features of the  ``deep Hilbert space'' (DHS). We study the out-of-time order correlator (OTOC) in the infinite-temperature ensemble of the full Hilbert space. We derive a phase-space representation of the DHS OTOC and show that the OTOC can have a super-exponential initial growth in the large $N$ limit, due to the fast dynamics in an unbounded phase space {(in finite systems, we observe numerically that the super-exponential growth ends precociously and gives way to a power-law one until saturation)}. By a similar mechanism, the Krylov complexity grows explosively. We also study the entanglement growth in a quantum quench from a DHS product state, i.e., one of non-aligned spins that resemble the DHS infinite-temperature ensemble with respect to the statistics of the collective spins. Using a field-theoretical method, We exactly calculate the entanglement entropy in the large $N$ limit. We show that, in the DHS, fast OTOC growth does not imply fast entanglement growth, in contrast to the Zurek-Paz relation derived in the TSS. 
\end{abstract}
\maketitle

\tableofcontents
\section{Introduction}\label{sec:intro}

 Recently, there has been much interest in many-body quantum systems with all-to-all interactions. Roughly speaking, two kinds of such systems are widely considered. The first is those with random coupling coefficients, as exemplified by the Sachdev-Ye-Kitaev (SYK) models~\cite{SY93,kitaev15,sykcomment,kitaevsuh,RevModPhys-syk}. These models are motivated by the possibility of ``simulating'' quantum gravity in the lab: their low-temperature regime are equivalent, via holography, to semiclassical gravitational systems involving a black hole. However, quantum systems with a semiclassical gravitational dual are notoriously hard to realize in the lab~\cite{maldacena2023simple,yaoSYK,kobrin2023comment,xu2020sparse,sparseSYK,jaferris,chen1028,lowrank}, in particular because of the random coupling coefficients. 

 \begin{figure}[h]
    \centering
    \includegraphics[width=0.96\columnwidth]{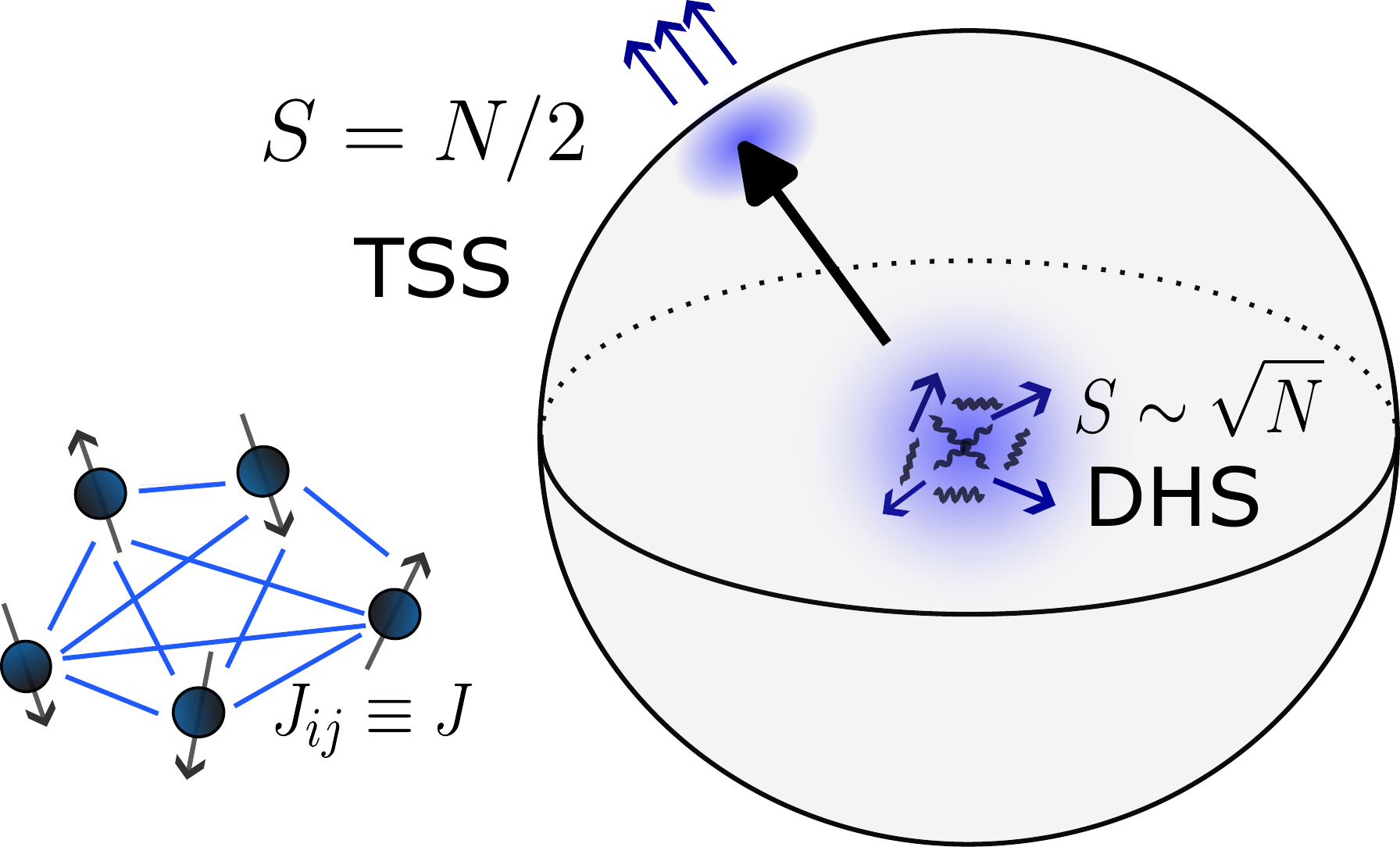}
    \caption{A cartoon of the many-body Hilbert space of a spin system with all-to-all interaction and \textit{uniform} coefficients. It is divided into conserved sectors. In the totally symmetric space (TSS), which is invariant under permutation of sites, the collective spin $ S = \sum_j {S}_j$ scales as the system size $S\sim N$.  In the large $N$ limit, a semiclassical description leads to a phase-space of a two-sphere. The vast majority of the states are in the deep Hilbert space (DHS), where the collective spin scales as $S \sim \sqrt{N}$. The phase space of the DHS is the interior the TSS-sphere and becomes non-compact in the $N\to\infty$ limit.
    } 
    \label{fig:schematic}
\end{figure}
 The second kind, which are more accessible experimentally~\cite{colciaghi2022einstein,albiez2005,leroux2010,davis2019,bollinger2016}, have \textit{uniform} coupling coefficients. For definiteness, consider a system of $N$ spin one-halves represented by local operators $S_1^a, \dots, S_N^a$, $a = x,y,z$, with uniform all-to-all interaction described by a Hamiltonian that only involves collective spin operators:
 \begin{align} \label{eq:HandSa}
 & H=  S  \, h(\{\mathcal{S}_a\})  \,,\, \mathcal{S}_a := \frac1{S} \sum_{i=1}^N S_i^a \,,\, S := N / 2 \,. 
 \end{align}
Here $h(\{\mathcal{S}_a\})$ denotes a polynomial of $\mathcal{S}_a$. When $N\to\infty$, such systems have a semiclassical limit in a non-holographic sense, provided we restrict ourselves to the \textit{totally symmetric space} (TSS). This is the subspace of states invariant under permutations of sites, and it is preserved by the dynamics. The semiclassical limit then follows from the fact that the TSS is a representation with spin $S = N/2$ of the SU(2) algebra formed by the collective spins, with a small \textit{effective} Planck constant $\hbar_{\text{TSS}} = 1/S$ (this is why the Hamiltonian \eqref{eq:HandSa} has an overall factor $ S=1/\hbar_{\text{TSS}}$, so that the time evolution operator is $e^{- i H t}$). Restricting to the TSS is a common practice, and also a reasonable thing to do as it is often convenient to prepare experimentally initial states in the TSS, such as a spin coherent state. It is worth noting that quantum dynamics with non-local interactions can be ``confined'' in the TSS to a good approximation even when the Hamiltonian deviates from the ideal all-to-all form~\eqref{eq:HandSa}, e.g., when the interaction has a (slow enough) power-law decay in distance~\cite{pappalardilerosePRR,pappalardi18}, or when the collective spin is deformed to $\propto \sum_i c_i S_i^a$~\cite{monika-ehud,Monika}. The dynamics inside the TSS is highly interesting from a number of perspectives, e.g., fast scrambling of quantum information~\cite{may-otoc-measure}, realization of macroscopic entanglement~\cite{colciaghi2022einstein,polzik01entanglement,gross10LMG}, applications to metrology~\cite{rev-metrology} and steeting~\cite{schrodinger_1935,steering-prl}, etc. For these reasons, much theoretical work has been devoted to various aspects of quantum dynamics in the TSS --- entanglement growth~\cite{bianchi18,hackl2018pra,lerose-pappalardi2020pra,pappalardilerosePRR}, the growth of out-of-time order correlators (OTOC)~\cite{Larkin1969QuasiclassicalMI,maldacena2016bound,cotler2018,swingle}, and Krylov complexity~\cite{hyp,Bhattacharjee_2022}, etc.

In this paper, we address the question: What is the nature of the quantum dynamics of all-to-all systems far away from the TSS? The TSS, of dimension $N+1$, occupies a shallow surface of the full Hilbert space of dimension $2^N$. In this regard, focusing on the TSS ignores the elephant in the room. To explore the genuine quantum many-body aspects of all-to-all models, we must dive in to the \textit{deep Hilbert space} (DHS). This question is also motivated by the need to relate the two kinds of all-to-all models~\cite{monika-ehud}. Those like SYK do not have an equivalent of a TSS (see however \cite{scaffidi19,haldar-2021,lowrank}); their quantum dynamics always explores an exponentially large Hilbert space, even at low energies. It turns out that the quantum dynamics of all-to-all systems in the DHS has several surprising features --- in particular, super-exponential growth of OTOCs and explosive growth of K-complexity --- which makes the DHS dynamics distinct from its counterparts in both SYK and TSS.  

\section{Overview}
\begin{table*}[]
    \centering
    \begin{tabular}{|c|c|c|c|}
     \hline
       &  Totally Symmetric Space & Deep Hilbert Space &  Section  \\ \hline 
     Collective Variable  &  $\mathcal{S}_a = \frac1{S} \sum_j S_j^a$   & $\mathbf{S}_a = \frac1{\sqrt{N}} \sum_j S_j^a$ & \\ 
     Hamiltonian & $H = S h (\{\mathcal{S}_a\})$  &  
      $H=\sqrt{N} h(\{\mathbf{S}_a\})$    & \\
      &   $1/N$ normalization & $1/\sqrt{N}$ normalization &\\ \hline
      $b_n$ (Lanczos coefficients)   &  $\alpha n$ & $ n^{3/2}$ &\ref{sec:k-review}, \ref{sec:k-dhs}  \\
      $K(t)$ (K-complexity)  &  $\exp({2\alpha t})$ & $(t_c-t)^{-2} $ & \ref{sec:k-review}, \ref{sec:k-dhs} \\ \hline 
      OTOC definition & $\mathrm{Tr}[\rho [\mathcal{S}_a(t),\mathcal{S}_b]^2  ]$   &
      $\sum_j \mathrm{Tr}[\rho [\mathbf{S}_a(t), S_j^b]^2  ]$  & \ref{sec:otoc-tss-dhs}
      \\ 
      OTOC growth  & $\exp({\lambda t})$ & $\exp({c t^2})$ & \ref{sec:super-exp-OTOC} \\ \hline
      Entanglement $S(t)$ & $t$ (saddle) or $\ln t$ & $\ln t$ (*) & \ref{sec:entanglement} \\ \hline 
     \end{tabular}
    \caption{Summary of the main results of this work, for the Euler top model with generic coupling constants. We display the asymptotic behavior of various quantities in the large $N$ limit. (*) Note that for the entanglement growth in the DHS and in the TSS, the correct normalization is the $1/N$ one.}
    \label{tab:summary}
\end{table*}
Let us start by defining the term deep Hilbert space (DHS). Consider the  collective spin variables, but with a different normalization [compare with \eqref{eq:HandSa}]:
\begin{equation}\label{eq:SaDHS}
\mathbf{S}_a := \frac1{\sqrt{N}} \sum_{i=1}^N S_i^a \,,
\end{equation}
which we shall refer to as the DHS collective spins. A state $\rho$ is in the DHS if the DHS collective spins have order one fluctuation:
\begin{equation} \label{eq:defDHS}
    \Tr[\rho f(\{ \mathbf{S}_a \}) ] = O(1) \,, 
\end{equation}
in the $N\to\infty$ limit. Note that a state in the TSS does \textit{not} satisfy \eqref{eq:defDHS} since the $\mathbf{S}_a$ have $O(\sqrt{N})$ fluctuations. 
A prototypical example of a DHS state, to which much of this work will be devoted, is the infinite-temperature ensemble of the full Hilbert space
\begin{equation}
    \rho_{\infty} := 2^{-N} \mathbf{I}, 
\end{equation}
There, $\mathbf{S}_a$ behave as Gaussian variables. Another example of a DHS state is given by product states of misaligned spin-$1/2$'s, see \eqref{eq:Psi0intro} below.

The DHS collective spins satisfy the SU(2) algebra with a different effective Planck constant: 
\begin{equation} \label{eq:DHSSU2}
[\mathbf{S}_x, \mathbf{S}_y] = \hbar_{\text{DHS}}\mathbf{S}_z  \,,\, \hbar_{\text{DHS}} = 1/\sqrt{N}  \,. \end{equation}
Therefore, the dynamics in the DHS has a parametrically different time scale compared to the TSS, which turns out to be the origin of many surprising phenomena in the DHS, as already mentioned. To explain this more precisely, we may consider an autocorrelation function such as
   \begin{equation} 
      G^a(t) := \Tr[ \rho_{\infty} \mathbf{S}_a(t) \mathbf{S}_a] \,\label{eq:Gt}
   \end{equation}
For it to have a well-defined $N\to\infty$ limit, we must normalize the all-to-all Hamiltonian differently, by using the DHS collective variables~\cite{MuellerLiu} and effective Planck constant $\hbar_{\text{DHS}}$ [compare to \eqref{eq:HandSa}]:
\begin{align} 
\label{eq:HDHS}
&  H = \sqrt{N} h(\{  \mathbf{S}_a \}) \,.
\end{align}
The same can be said about OTOCs, provided we adopt an appropriate  definition adapted to the DHS, see Section~\ref{sec:otoc-tss-dhs} below. In what follows, we shall refer to the normalization of \eqref{eq:HandSa} the $1/N$ normalization and \eqref{eq:HDHS} the $1/\sqrt{N}$ normalization. 

The issue of normalization is not a formality, but has a tangible physical consequence: the terms in the Hamiltonian that are nonlinear in $S_j^a$ lead to a dynamics that is parametrically faster in the TSS than in the DHS. Let us preview how this is responsible for the super-exponential OTOC growth (see Section~\ref{sec:otoc}). The OTOC in the DHS admits a phase-space representation. The phase space is $\mathbb{R}^3$ in the thermodynamic limit, parametrized by the DHS collective variables $\mathbf{S}_a$. In a finite system, the phase space is cut off and becomes a ball whose boundary sphere is the classical phase space of the TSS~(see Figure~\ref{fig:schematic} for a sketch). Thus, as one moves away from the origin and towards the TSS, the dynamics becomes faster. As we shall see, this faster dynamics dominates the OTOC and makes it grow super-exponentially in paradigmatic all-to-all models. 

{{red}Interestingly, despite the super-exponential OTOC growth, the all-to-all models in the DHS are \textit{not} fast scramblers. Indeed, we computed the OTOC with an exact numerical method that allows to access large finite systems (up to $N \sim 100$). As a result, we find that the super-exponential growth ends precociously, and is replaced by a slower power-law growth until the final saturation. The saturation time $t_S$ scales as a power law of $N$ (whereas by definition~\cite{Sekino_2008}, a fast scrambler has $t_S \sim \ln N$). This is an unusual finite-size effect, which awaits an analytic understanding; see however a heuristic discussion in Section~\ref{sec:finiteNOTOC}.}

A similar mechanism is behind the explosive growth of K-complexity, which is a measure of operator growth proposed by some of us~\cite{hyp}. It is simpler than the OTOC in that it only requires the knowledge of the autocorrelation function. In the DHS, the latter also has a phase space representation~(Section~\ref{sec:semiclassic-Gt}). The acceleration of the dynamics away from the phase space origin gives rise to an anomalously fat tail in the spectral density, which leads to the fast K-complexity growth~\cite{Lubinsky1987,viswanath2008recursion,pettifor2012recursion}. In fact, a closely related result (in terms of Lanczos coefficients obtained with the recursion method) was first observed in Ref.~\cite{MuellerLiu}, in the context of classical all-to-all XYZ model. Here, we will explain their observation, extend it to quantum models, and interpret the result in terms of K-complexity growth (Section~\ref{sec:k-review} and \ref{sec:k-dhs}). 

The last part (Section~\ref{sec:entanglement}) of this work is devoted to the entanglement growth following a quantum quench from a DHS product state. We were particularly motivated by the relation between scrambling/chaos and entanglement entropy growth~\cite{hosur-qi,may-otoc-entanglement,may-otoc-entanglement-2}. In semiclassical systems, such a relation was put forward by Zurek and Paz~\cite{zurek-paz,zurek-paz2}, and has since been rather well established, see e.g.,~\cite{sarkar98,furuya98-entanglementchaos,gong03}. Roughly speaking, the entanglement growth rate is given by the sum of positive Lyapunov exponents (of the linearized dynamics around the classical trajectory), which also govern the OTOC growth~\cite{pappalardi18}.  Does such a relation exist in the DHS, so that the super-exponential OTOC growth gives rise to a super-linear entanglement growth? 

To address this question, we consider a quantum quench starting from a product state
\begin{equation}
        \vert \Psi_0 \rangle = \prod_{j=1}^N \vert s_j \rangle  \label{eq:Psi0intro}
\end{equation}
where $\{s_j\}_{j=1}^N$ is a set of distinct spin-$1/2$ states forming a smooth distribution on the Bloch sphere (as $N\to\infty$) whose center of mass is at the origin, $\sum_j s_j = 0$. We shall show that such a state is in the DHS, by the definition \eqref{eq:defDHS}. In fact, it is similar to the DHS infinite-temperature state $\rho_\infty$ in that the DHS collective spins have also Gaussian statistics. Thus, the OTOC will have a similar super-exponential growth. 

However, the dynamics of entanglement appears to be completely unrelated to OTOC. First, they have distinct time scales: to obtain a well-defined $N\to\infty$ limit of the bipartite entanglement entropy growth, the $1/N$ normalization of the Hamiltonian \eqref{eq:HandSa} [\textit{not} the $1/\sqrt{N}$ one \eqref{eq:HDHS}!] is the appropriate one to obtain a large $N$ limit. We will show this and calculate exactly the (Renyi) entanglement entropy in the large $N$ limit, as the semiclassical expansion (one-loop determinant) of a path integral. The semiclassical picture that emerges is unrelated with the phase space picture of the OTOC. In particular, it predicts a logarithmic entanglement entropy growth in situations where the OTOC grow super-exponentially. Finally, we find numerically that the entanglement growth saturates at $ O(\ln N)$, not a volume law. We conclude that the DHS state \eqref{eq:Psi0intro} resembles a TSS state in many regards. This suggests the existence of a ``depth hierarchy'' in the Hilbert space of which we have merely scratched the surface. 

The main results of this work are summarized in Table~\ref{tab:summary}, where we also provide the relevant sections. 

\section{Autocorrelation function and K-complexity}
In order to study the growth of K-complexity in the DHS, we first develop a phase space representation of the autocorrelation function (Section~\ref{sec:semiclassic-Gt}). (This representation will also be useful for the study of OTOCs in Section~\ref{sec:otoc}).
Section~\ref{sec:k-review} reviews the basics of K-complexity needed to appreciate the new results in the DHS, reported in Section~\ref{sec:k-dhs}. 


\subsection{Phase space representation of the autocorrelation function}\label{sec:semiclassic-Gt}
We consider the autocorrelation function \eqref{eq:Gt} in an all-to-all model with the $1/\sqrt{N}$ normalization, extending Ref.~\cite{MuellerLiu} which focused on classical all-to-all models. Although they considered the specific example of the Euler top (see below), a main result of Ref.~\cite{MuellerLiu} can be stated for general quantum spin models as follows: 

\noindent\textbf{Phase space representation of $G^a(t)$}. In the large $N$ limit, the autocorrelation function $G^a(t)$ as defined in \eqref{eq:Gt} admits a phase space representation:
\begin{equation} \label{eq:Gtres}
    \lim_{N\to\infty}G^a(t) =  \left< s_a(t) s_a \right> \,.
\end{equation}
Here, $\{s_a: a = x,y,z\}$ are classical variables, and $\left< [\dots] \right>$ denotes a phase space average 
\begin{equation}\label{eq:phasespacedef}
    \left< f(\{s_a\}) \right> = \int  f( \{s_a\} ) \prod_b e^{-2 s_b^2 } \frac{\mathrm{d} s_b }{\sqrt{\pi / 2}} \,.
\end{equation}
Finally, $s_a(t)$ is a function of $\{s_a\}$ determined by the equation of motions:
\begin{equation}
    \dot{s}_a(t) = \{s_a(t), h  \}_{\text{P.B.}} \,,\, s_a(0) = s_a \,. \label{eq:dsadt}
\end{equation}
where $\{\cdot, \cdot\}_{\text{P.B.}}$ is the SU(2) Poisson bracket, with e.g. $\{s_x, s_y\}_{\text{P.B.}} = s_z$, and where $h = h(\{s_a\})$ is the same as in \eqref{eq:HDHS}. 
 For example, the Euler top (also known as the XYZ model) has the Hamiltonian 
\begin{equation}
    H =  \frac{\sqrt{N}}2 (J_x \mathbf{S}_x \mathbf{S}_x + J_y \mathbf{S}_y \mathbf{S}_y + J_z \mathbf{S}_z \mathbf{S}_z ) \,.
\end{equation}
Then $h = \frac12 \sum_a J_a s_a^2$, and the equation of motions~\eqref{eq:dsadt} are 
\begin{align}
    \dot{s}_x = (J_y - J_z) s_z s_y \,,\, 
\end{align}
and its cyclic permutations. 

We can show the proposition~\eqref{eq:Gtres} by a rather standard semiclassical argument. This consists of two independent observations, which will be useful in Section~\ref{sec:otoc} below. Let us review them in turn. 

First, time evolution of the operators is classical. This is because the collective variables $\mathbf{S}_a$ satisfy an SU(2) commutator algebra 
with a small effective Planck constant~\eqref{eq:DHSSU2}, i.e., they almost commute. Thus, in the $N\to \infty$ limit, we have the familiar quantum-classical correspondence: operators become classical functions on the phase space, $\mathbf{S}_a \to s_a $, and commutators become Poisson brackets $[\cdot, \cdot] \to  i \hbar_{\text{DHS}} \{\cdot, \cdot \}_{\text{P.B.}}$. When considering time evolution, the effective Planck constant is cancelled by the $\sqrt{N}$ factor in \eqref{eq:HDHS}, giving rise to \eqref{eq:dsadt}. All this is similar to the SU(2) algebra of the TSS collective variables, where the effective Planck constant is $1/S$.

Second, Eq.~\eqref{eq:phasespacedef} means that the average over the infinite-temperature ensemble $\rho_{\infty}$ corresponds, as $N\to\infty$, to a phase space average over $\mathbb{R}^3$ in which the $\mathbf{S}_a$'s behave as independent Gaussian variables with vanishing mean and standard deviation $1/2$. To see this, one may compute the generating function and show that (see Appendix \ref{app:traces})
\begin{align}\label{eq:Mua}
   M(\{u_a\}) := \Tr[\rho_{\infty} \prod_{a} \exp( u_a \mathbf{S}_a) ]  = \prod_a e^{u_a^2 / 8} \,,
\end{align}
which characterizes the Gaussian distribution we just described. The integration measure $\prod_a e^{-2s_a^2} \mathrm{d} s_a / \sqrt{\pi/2}$ of \eqref{eq:phasespacedef} is the probability density function of this distribution.  

Combining the two observations, we may derive the claim \eqref{eq:Gtres} immediately. In the rest of the section, we will use \eqref{eq:Gtres} to study K-complexity growth in the DHS.

\subsection{Krylov complexity: generality}\label{sec:k-review}

Let us briefly review the general K-complexity approach to many-body quantum chaos~\cite{hyp} before applying it to the DHS of all-to-all systems in the next subsection.
The general idea is the following. Given the data of (i) a Hamiltonian $H$, (ii) an Hermitian operator $O$, and (iii) an inner product on the operator space, which we shall take to be the infinite-temperature one: 
\begin{equation}\label{eq:infiniteTinner}
    (A | B ) := \mathrm{Tr}[\rho_{\infty} A^\dagger B] \,, 
\end{equation}
one may apply the Gram-Schmidt procedure to the sequence of operators $\{O, [H, O], [H, [H, O]], \dots, \}$ generated in the Heisenberg time evolution of $O$ under $H$. The resulting orthonormal sequence $\{O_n\}_{n=0}^\infty$ has the interesting property of tri-diagonalizing the action of $[H, \cdots]$ (known as the Liouvillian). Namely, there exists a set of positive Lanczos coefficients $\{b_n\}_{n=1}^\infty$ such that 
\begin{equation} \label{eq:tri-diag}
    [H, O_n] = b_{n+1} O_{n+1} + b_{n} O_{n-1} \,,\, n = 0, 1, 2, \dots \,,
\end{equation}
where $O_{-1} = 0$ and $b_{0} = 0$ by convention. In fact, both $b_n$ and $O_n$ can be found by the well-known Lanczos algorithm (which is an optimization of the Gram-Schmidt procedure). Physically, we may interpret \eqref{eq:tri-diag} as mapping the time evolution of $O(t)$ in the space of many-body operators to a single-particle quantum mechanics problem on a semi-infinite chain. The wave-function, defined as the expansion of $O(t)$ in the basis $\{O_n\}$
\begin{equation}
     \varphi_n(t) = i^n (O_n |O(t)) 
\end{equation}
satisfies a Schr\"odinger equation:
\begin{align}\label{eq:dirac}
   \dot{\varphi}_n(t) = b_{n} \varphi_{n-1} - b_{n+1} \varphi_{n+1}.
\end{align} 
In terms of the quantum mechanics problem, the autocorrelation $G(t)$ is simply the amplitude of the particle returning to the origin after time $t$: this is the basis of the recursion method, a well-established tool in linear response calculations~\cite{viswanath2008recursion} (see \cite{auerbach2018hall} for a recent application). Meanwhile, as advocated in \cite{hyp}, to make connection with quantum chaos, one should rather focus on the spreading of the wavefunction along the semi-infinite chain. Indeed, the K-complexity is defined as the expected position of the operator-wavefunction:
\begin{equation}
    K(t) = \sum_n n |\varphi_n(t)|^2 \,.
\end{equation}
Since the operators $O_n$ are more non-local in general,  $K(t)$ is a measure of complexity growth of $O(t)$. It is closely related to OTOC and operator size, especially in SYK models~\cite{hyp,openquantumsystems3}. 

Although related to OTOCs, $K(t)$ is completely determined by the two-point function $G(t)$, through its Fourier transform (the spectral function $\rho(\omega)$), and the Lanczos coefficients. In particular, some of us conjectured that in generic chaotic systems, the spectral function has an exponential tail, the Lanczos coefficients grow linearly in $n$, and the K-complexity increases exponentially in $t$~\cite{hyp,Cao_2021}:
\begin{equation}
    \rho(\omega) \sim e^{-|\omega| / \omega_0 } \,,\, b_n \sim \alpha n \,,\, K(t) \sim e^{2\alpha t} \,. \label{eq:UOGP}
\end{equation}
The latter growth rate can be quantitatively compared to the OTOC Lyapunov exponent in SYK models.

\subsection{K-complexity explosion in the DHS}\label{sec:k-dhs}
Although the universal operator growth hypothesis summarized in~\eqref{eq:UOGP} was observed to hold in a broad variety of systems, it is not expected to apply to all-to-all models with the $1/\sqrt{N}$ normalization.
Indeed, the operator growth hypothesis assumes an extensive many-body bandwidth, which is not the case under the $1/\sqrt{N}$ normalization since the ground state (or highest state) energy is super-extensive due to states in the TSS\footnote{Unless the Hamiltonian is linear in $\mathbf{S}_a$}.
For example, the energy of TSS states scales as $\sim N^{3/2}$ in a quadratic Hamiltonian like the Euler top. Thanks to a super-extensive bandwidth, the spectral function can have a fatter-than-exponential tail, which means the Lanczos coefficients $b_n$ can grow faster than linearly, and that the K-complexity $K(t)$ can grow faster than exponentially. This is indeed observed numerically in Ref.~\cite{MuellerLiu}, in which the authors reported
\begin{equation} \label{eq:Eulerb_n}
   \rho(\omega) \sim e^{- (|\omega| / \omega_0)^{2/3} } \,,\,  b_n \sim n^{3/2}  
\end{equation}
for the Euler top with generic (unequal) couplings. 

We now provide an analytical explanation for \eqref{eq:Eulerb_n}. To do this, recall from~\cite{MuellerLiu} that the total classical spin $ \sum_a s_a^2 $ is conserved by the classical dynamics~\eqref{eq:dsadt}. Therefore, the phase space average \eqref{eq:Gtres} can be decomposed as an integral over $r$:
\begin{equation}
    G^a(t) = C \int_{0}^\infty \left< s_a(t) s_a \right>_r  e^{-2 r^2} r^2 \mathrm{d} r
\end{equation}
where the $N\to\infty$ limit is tacitly assumed, $C = 8\sqrt{2/\pi}$ is an unimportant constant and $\left< [\dots] \right>_r$ is a phase average like \eqref{eq:phasespacedef}, except that it is on a sphere of radius $r$ (with area normalized to $1$):
\begin{equation}
    \left< f(\{s_a\}) \right>_r = \frac{1}{4\pi r^2} \int_{\sum_a s_a^2 = r^2}  f(\{s_a\}) \,.
\end{equation}
Similarly, we have for the spectral function
\begin{equation}
    \rho(\omega) =C \int  \rho_r(\omega) e^{-2 r^2} r^2 \mathrm{d} r \label{eq:rhoasintegral}
\end{equation}
where $\rho_r(\omega)$ is the Fourier transform of $\left< s_a(t) s_a \right>_r$. Now, the key observation is that $\left< s_a(t) s_a \right>_r $ and $\rho_r(\omega)$ are respectively the autocorrelation and spectral function of the Euler top model in the TSS infinite-temperature ensemble, with rescaled couplings $J_a \to r J_a$. Adapting phase-space methods in \cite{Bhattacharjee_2022}, one may show that $\rho_{r=1}(\omega) \sim e^{-|\omega|/ \omega_0}$, in agreement with the general conjecture~\eqref{eq:UOGP}. Thus, we have $\rho_r(\omega) \sim  e^{-|\omega| / r\omega_0} / r $ by time rescaling. Plugging this into \eqref{eq:rhoasintegral} and taking a saddle point approximation for large $\omega$, we have
\begin{equation}
     \rho(\omega) \sim \int  e^{-|\omega| / (r \omega_0)  - 2 r^2} r \mathrm{d} r \sim e^{- c_1 (|\omega| / \omega_0)^{2/3}} \,, \label{eq:rho_Euler}
\end{equation}
where $c_1=3 / 2^{1/3}$ and where we omitted subleading, power-law in $\omega$, terms. This is exactly the spectral function tail of \eqref{eq:Eulerb_n}. Then, the Lanczos coefficients asymptotics in ~\eqref{eq:Eulerb_n} follows from a known general dictionary~\cite{viswanath2008recursion,avdoshkin} (see Appendix~\ref{app:lanczos}.)  

\begin{figure}
    \centering
    \includegraphics[width=\columnwidth]{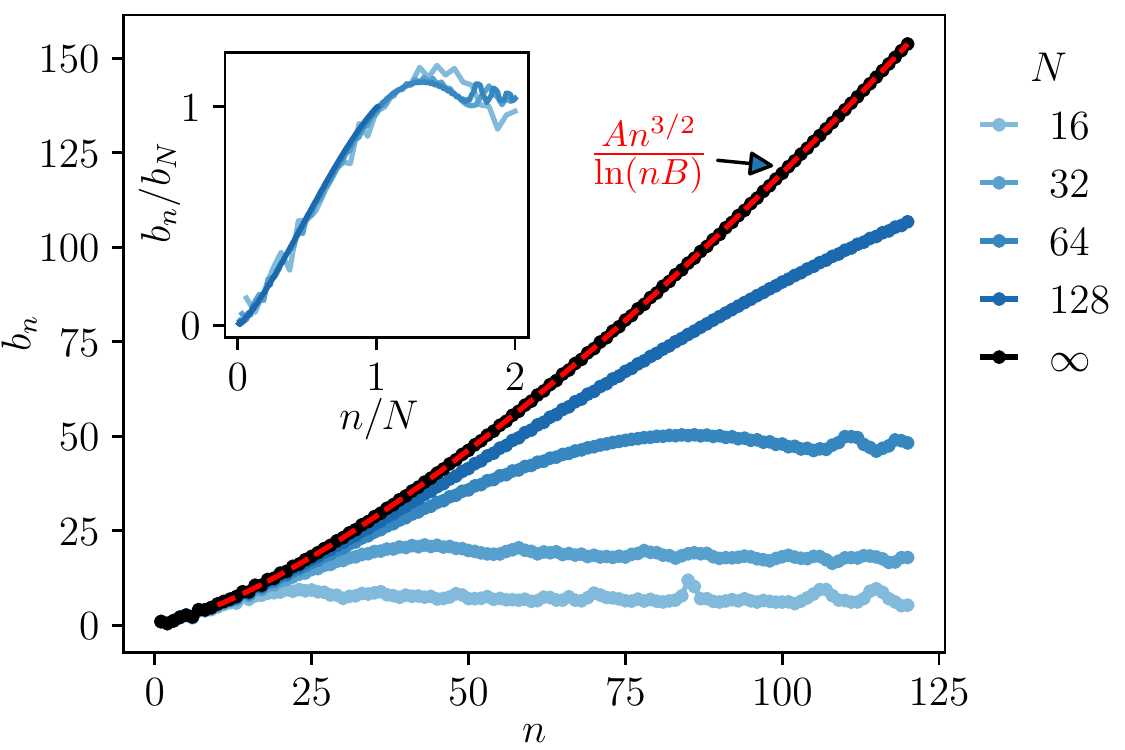}
    \caption{Lanzcos coefficients $b_n$ of the operator $\mathbf{S}_z = N^{-1/2} \sum_j S_j^z$ in the LMG model $H = \sum_{j} S_j^x + N^{-1/2} J \sum_{jk} S_j^z S_k^z / 2$~\eqref{eq:HLMG} with $J=1$, and with respect to the infinite-temperature ensemble $\rho_{\infty}$. \textbf{Main}. we plot $b_n$ for a few system sizes, as well as in the thermodynamic limit. The latter is fitted by the prediction $b_n = A n^{3/2} / \ln (n B)$~\eqref{eq:bnLMG}; the fit parameters are $A = 0.95$ and $B=27.2$. \textbf{Inset}. Data of different finite system sizes are collapsed by plotting $b_n / b_N$ against $n/N$. This confirms that the growth of $b_n$ saturates at $n_{\text{sat}} \sim N$~\eqref{eq:bnfiniteN}.  }
    \label{fig:dhs-lanczos}
\end{figure}

The upshot of the analysis is that, the anomalously fat tail in the spectral function results from contributions of large $r$, where the dynamics is faster. This mechanism applies qualitatively to generic all-to-all models with $1/\sqrt{N}$ normalization, while the precise exponent of $|\omega|/\omega_0$ and that of $n$ in \eqref{eq:Eulerb_n} are model dependent. To illustrate, let us consider another paradigmatic example, the Lipkin-Meshkov-Glick (LMG) model~\cite{GLICK1965211,*MESHKOV1965199,*LIPKIN1965188}:
\begin{equation} \label{eq:HLMG}
    H = \sqrt{N} \left( \mathbf{S}_x +  \frac{J}2   \mathbf{S}_z  \mathbf{S}_z  \right) \,.
\end{equation}
Unlike the Euler top, the LMG Hamiltonian is not a homogeneous polynomial of $\mathbf{S}_a$. So, while \eqref{eq:rhoasintegral} still applies, the TSS classical dynamics with different $r$'s are not related by a simple time rescaling. However, we can still show that $ \rho_r(\omega) \sim e^{-|\omega|/ \omega_0(r)}$ with $\omega_0(r) \sim r / \ln r$ for large $r$, using the exact result of \cite{Bhattacharjee_2022} (see Appendix~\ref{app:LMG}). Thus, \eqref{eq:Eulerb_n} also holds for the LMG model, up to a log-correction:
\begin{equation}\label{eq:bnLMG}
    b_n \sim \frac{ n^{\frac32}}{\ln n}  \quad \text{(LMG)}\,.
\end{equation}
We tested this prediction numerically, see Figure~\ref{fig:dhs-lanczos}.  In general, we expect that a $k$-local all-to-all Hamiltonian (i.e., a degree-$k$ polynomial of $\mathbf{S}_a$'s) leads to the following:
\begin{equation}
    \rho(\omega) \sim \exp(-c (|\omega| / \omega_0)^{\frac{2}{k+1}} ) \,,\, b_n \sim n^{\frac{k+1}2} \,.
\end{equation}

The consequence of a super-linear growth of $b_n$ on the K-complexity is rather explosive: formally, $K(t)$ diverges to $+\infty$ in finite time. Indeed, the continuum limit of \eqref{eq:dirac}, 
\begin{equation}
(\partial_t + 2  b_n \partial_n) \varphi = 0,
\end{equation}
has characteristic curves $ n = (t - t_c)^{-1/(a-1)} $ for any super-linear growth $b_n \sim n^{a}$, $a > 1$. This implies that a nonzero weight of the wavefunction is transported to $n\to \infty$, and thus $K(t) \to \infty$, as $t\to t_c$ for some finite $t_c$. 

The finite-time explosion of the K-complexity may seem non-physical. Indeed, it is an artefact of the $N\to\infty$ limit. For a system with finite (but large) $N$, we claim that the growth of the Lanczos coefficients as in \eqref{eq:Eulerb_n} saturates in the following way:
\begin{equation} \label{eq:bnfiniteN}
    b_n \sim \begin{cases} 
    n^a  & n \lesssim n_{\text{sat.}} \sim N \,,\\
    N^a  & n \gtrsim n_{\text{sat.}} \,,
    \end{cases} \,,\, a = \frac{k+1}2
\end{equation}
for a generic $k$-local Hamiltonian; in Figure~\ref{fig:dhs-lanczos} (inset), we showcase this behavior using the LMG model. The saturation of the $b_n$ regularizes the finite-time divergence of the K-complexity. Instead, the latter takes a $O(1)$ amount of time (as $N\to\infty$) to reach $ \sim n_{\text{sat.}}$, and then increases linearly with velocity $\sim N^a$ until reaching the end of the Krylov space. One simple way to understand the saturation scale of \eqref{eq:bnfiniteN} is the following: the operators $\mathcal{O}_n$ generated by $n$ iterations of the Liouvillian is almost $[k n + O(1)]$-body, if the Hamiltonian is $k$-local. Therefore, the operator starts to fill a significant portion of the available space only as $n\sim N$. Alternatively, one may recall that $b_n$ is bounded by the norm of the Liouvillian, which is in turn bounded by the bandwidth of the many-body energy spectrum, which scales as $N^a$.   

To summarize this section, we revisited the work of M\"uller and Liu~\cite{MuellerLiu}, which revealed for the first time (to our knowledge) peculiar features of quantum dynamics in the DHS. We explained analytically their numerical observation, and interpreted it in terms of an explosive growth of the K-complexity. Such a behavior, forbidden in ``usual'' systems, is a first manifestation of the weirdness of the DHS. Similarly strange behavior appears for OTOCs, another measure of operator growth, as we shall see next.

\section{Scrambling in the deep Hilbert space}\label{sec:otoc}
\subsection{Out-of-time ordered correlators: TSS vs DHS}\label{sec:otoc-tss-dhs}
In this section, we study the growth of OTOCs in the DHS of all-to-all models. More precisely, we will consider squared commutators of the following form:
\begin{equation} \label{eq:OTOCintro}
      \mathbf{C}(t) := \frac12  \sum_{a, j} \mathrm{Tr}\left( \rho_{\infty} [\mathbf{O}(t), S_j^a] [\mathbf{O}(t), S_j^a]^\dagger \right)  \,,\,
\end{equation}
where $\mathbf{O}$ is a function of the collective variables:
\begin{equation}
      \mathbf{O} = f(\{\mathbf{S}_a\}) \,.
\end{equation}
The sum over $a$ and the factor $1/2$ are included in order to streamline the connection with operator size, see below. 

Before proceeding any further, it is useful to compare \eqref{eq:OTOCintro} with OTOCs in TSS, which are of a distinct form:
\begin{equation}\label{eq:OTOCclassical}
         \mathbf{C}_{\text{TSS}}(t) :=  \sum_a \mathrm{Tr}\left( \rho [ \mathcal{O}(t),  \mathcal{S}_a ] [\mathcal{O}(t), \mathcal{S}_a]^\dagger \right)  \,.
  \end{equation}
Here, $\mathcal{O}$ is a function of $\mathcal{S}_a$, and $\rho$ is a density matrix whose eigenstates all belong in the TSS. For example, it can be the infinite-temperature state of the TSS, (i.e. $P_\text{TSS}/(2S+1)$ with $P_\text{TSS}$ the projector onto the TSS), or a pure coherent state $ (|s\rangle \langle s | )^{\otimes N}$.
However, the most crucial feature of $\mathbf{C}_{\text{TSS}}(t)$ is that it involves only collective variables. Therefore, in the large-$N$ limit, it lends itself readily to a semiclassical analysis, which is by now well-understood. As a result, one finds that classical chaos~\cite{rozenbaum17,dicke,cotler2018} or local instability~\cite{richter19,pilatowsky2019positive,scramblingvchaos,ExpOTOC1,saddledom1,rozenbaum19,steinhuber2023dynamical} can result in an exponential growth $\mathbf{C}_{\text{TSS}}(t) \sim e^{\lambda_L t}$ which saturates at the Ehrenfest time $t_E \sim \lambda_L^{-1} \ln N$ due to interference effects~\cite{richter}.

By contrast, the DHS OTOC \eqref{eq:OTOCintro} involves single-site operators $S_j^a$ \textit{individually}, not inside a collective variable. Indeed, the sum over sites in \eqref{eq:OTOCintro} is \textit{outside} the trace; due to the equivalence between sites, one may well replace it by a factor $N$:
\begin{equation}
     \mathbf{C}(t) =  \frac{N}2 \sum_{a} \mathrm{Tr}\left( \rho_{\infty} [\mathbf{O}(t), S_1^a] [\mathbf{O}(t), S_1^a]^\dagger \right)  \,. \label{eq:OTOC-alternative}
\end{equation}
Therefore, {naively}, a semiclassical analysis similar to that applied to the autocorrelation (see Section~\ref{sec:semiclassic-Gt}) does not apply here. We shall see however in the next section that it is possible to write a phase space representation for \eqref{eq:OTOCintro}. In order to do that, we shall recall a standard way of looking at \eqref{eq:OTOCintro}, that is, as measuring \textit{operator size}. Indeed, writing $O(t)$ as a linear combination of Pauli strings (here $\sigma_i^a = 2 S_i^a$ are local Pauli operators)
\begin{equation}
    \mathbf{O}(t) = \sum_{s} \sum_{j_1 < \dots< j_s} \sum_{a_1, \dots, a_s} c^{a_1, \dots, a_s}_{j_1, \dots, j_s}(t) \, \sigma_{j_1}^{a_1}\dots \sigma_{j_s}^{a_s} \,,
\end{equation}
where $s$ is the Pauli string length,
one may check that 
\begin{equation}
       \mathbf{C}(t) = \sum s \left|c^{a_1, \dots, a_s}_{i_1, \dots, i_s}(t) \right|^2 \,.
\end{equation}
In other words, the OTOC $  \mathbf{C}(t) $ measures the average size of the Pauli strings contained in $\mathbf{O}(t)$. This makes it a \textit{bona fide} measure of quantum information scrambling. By colloquial definition, the information carried by $\mathbf{O}(t)$ is scrambled if $\mathbf{O}(t)$ is dominated by highly many-body operators, that is, if it has large operator size. The operator-size interpretation of OTOCs is well-known,  especially in the SYK context~\cite{roberts-sykop,QiStreicher,qi2019measuring}, and is related to a number of interesting notions such as teleportation and size winding~\cite{teleportation1,teleportation2,teleportation3}.

Superficially, the DHS OTOC defined in \eqref{eq:OTOCintro} or \eqref{eq:OTOC-alternative} seems different from the standard definition
\begin{equation}\label{eq:standardOTOC}
    \mathrm{Tr}(\rho [S_i^a(t) , S_j^b]  [S_i^a(t),S_j^b]^\dagger) \,,
\end{equation}
namely, our operator under evolution $\mathbf{O}$ is a sum over local terms, while here $S_i^a$ is a single term. This is however a minor difference. Indeed, with a uniform all-to-all Hamiltonian, $S_i(t)$ will be a function of the collective spin variables $\mathbf{S}_a$ and local spin operators acting on site $i$. More precisely, in the large $N$ limit, one can show that (see Appendix~\ref{app:singlespin})
\begin{equation}\label{eq:Siat}
    S_i^a(t) = \sum_{b=x,y,z} S_i^b \,  \partial_{\mathbf{S}_b}f( \{\mathbf{S}_a\}, t)  
\end{equation}
where $f(t)$ evolves under the classical dynamics $\partial_t f = \{f, h\}_{\text{P.B.}}$ with initial condition $f(t=0) = s_a$.
Now, intuitively, the OTOC/operator size growth is dominated by the collective variables; the factors $S_i^b$ can change the operator size by at most one, so has a minor effect. As we shall show below, the growth of the OTOC~\eqref{eq:OTOCintro} is essentially equivalent to $f(t)$ becoming a fast-varying function on the phase space. When that happens, its derivatives $\partial_{s_b} f$ will be fast-varying as well. Hence, our approach below can be readily adopted to the OTOC~\eqref{eq:standardOTOC} (at the price of becoming more cumbersome), which will have the same qualitative behavior as \eqref{eq:OTOCintro}, modulo a $1/N$ prefactor. In what follows, we shall focus on the OTOC definition \eqref{eq:OTOCintro} for clarity.


\subsection{Phase space representation of OTOC}
We now derive a phase space representation of the OTOC \eqref{eq:OTOCintro}, see \eqref{eq:OTOCphase-final} below, starting from the operator-size formulation we just obtained. The result will be reminiscent of the phase-space representation of the OTOC in the TSS, see \eqref{eq:OTOCphase-TSS} below.

To do this, it is helpful to fully embrace the formalism of ``operator quantum mechanics'': i.e., we view the operator $\vert \mathbf{O}(t) )$ as a state living in the Hilbert space endowed with the inner product \eqref{eq:infiniteTinner}. Then the OTOC can be written as the expectation with respect to a super-operator $\mathbb{Q}$ measuring operator size:
\begin{align}
   &  \mathbf{C}(t)  =  \left( \mathbf{O}(t) | \mathbb{Q} |   \mathbf{O}(t) \right) \,,\,  \label{eq:Ct-Superop}\\
    &  \mathbb{Q} \vert \sigma_{i_1}^{a_1}\dots  \sigma_{i_s}^{a_s}  ) := 
    s \vert \sigma_{i_1}^{a_1}\dots \sigma_{i_s}^{a_s}  ) \,. \label{eq:Qdef}
\end{align}
Since the Pauli strings form an orthonormal basis of the operator Hilbert space, \eqref{eq:Qdef} defines the super-operator completely. Now, $| \mathbf{O}(t) )$ lives in the space of symmetric operators (invariant under site permutations). An orthonormal basis of this subspace are thus made of symmetrized sums of  Pauli strings. They are specified by the numbers $\ell, m, n$ of each type of Pauli's ($\ell + m + n \le N$), and defined as follows:
\begin{align}\label{eq:ellmndef}
    &| \ell, m, n )   \\ := &
    \binom{N}{\ell,m,n}^{-\frac12} \sum^*_{\substack{i_1 < \dots < i_\ell \\ 
    j_1 < \dots < j_m \\ k_1 < \dots k_n
    }} |\sigma_{i_1}^x  \dots _{i_\ell}^x \sigma_{j_1}^y  \dots \sigma_{i_m}^y  
     \sigma_{k_1}^z  \dots \sigma_{k_n}^z ) \,,\nonumber 
\end{align}
where the sum $\sum^*$ is over indices such that $\{i_1, \dots, i_\ell\}$, $\{j_1, \dots, j_m\} $, 
$\{k_1, \dots, k_n\} $ are mutually disjoint. There are thus $ \binom{N}{\ell,m,n} $ terms, hence the normalization factor. By definition, $| \ell, m, n )$ has fixed operator size, so 
\begin{equation} \label{eq:QonHermite}
   \mathbb{Q} | \ell, m, n )  = (\ell + m + n)  | \ell, m, n ) \,. 
\end{equation}

Next, we express $ | \ell, m, n )  $ in terms of the collective variables $\mathbf{S}_a$. Note that,  $  | \ell, m, n )$ are \textit{in general} not proportional to $\mathbf{S}_x^\ell \mathbf{S}_y^m \mathbf{S}_z^n $. For example, we can check explicitly that  $ \mathbf{S}_x^2 = \frac14 |0,0,0) + \frac{\sqrt{2}}{4} | 2, 0, 0 ) $. In finite systems, the relation between $  | \ell, m, n ) $ and $\mathbf{S}_a$ is rather involved. However, in the large $N$ limit, we have a simple result:
\begin{equation} \label{eq:ellmnisHermite}
     | \ell, m, n ) \stackrel{N\to\infty}= |  \chi_{\ell}(\mathbf{S}_x) \chi_{m}(\mathbf{S}_y) \chi_{m}(\mathbf{S}_z) )  
\end{equation}
where $\chi_n(x)$ are Hermite polynomials which we define as satisfying the following orthonormal relations:
\begin{equation} \label{eq:hermiteOrthonormal}
\int \chi_m(x) \chi_n(x) e^{-2x^2} \frac{ \mathrm{d} x}{\sqrt{\pi/2}} = \delta_{mn} \,.
\end{equation}
Eq.~\eqref{eq:ellmnisHermite} is shown in Appendices \ref{app:numerics} and \ref{app:hermite}. The gist of the proof can be understood by considering the operators $\vert \ell, 0,0 )$. They can be obtained by applying Gram-Schmidt to the sequence $ (\mathbf{S}_x^\ell)$, since $\mathbf{S}_x^\ell$ is a linear combination of $| \ell,0,0 )$, as well as $ \vert \ell', 0,0 )$ with $\ell' < \ell$. Now, recall that in the $N\to\infty$ limit, $\mathbf{S}_x$ behaves as a centered Gaussian of variance $1/4$, so that  
$$ ( f(\mathbf{S}_x) | g(\mathbf{S}_x)) = \int \overline{f(x)} g(x)  e^{-2x^2} \frac{ \mathrm{d} x}{\sqrt{\pi/2}} \,. $$
Comparing to \eqref{eq:hermiteOrthonormal}, we see that the Gram-Schmidt process will yield nothing but the Hermite polynomials of $\mathbf{S}_x$, and hence $ \vert \ell, 0,0 \rangle = \chi_\ell(\mathbf{S}_x)$. It remains to make sure that the ``interference'' between different Pauli species is vanishing at large $N$, as we do in Appendices \ref{app:numerics} and \ref{app:hermite}.

As a consequence of \eqref{eq:QonHermite} and \eqref{eq:ellmnisHermite}, we can express the action of $ \mathbb{Q}$ as a differential operator acting on the classical variables $\{s_a\}$. This is possible thanks to the differential equation satisfied by the Hermite polynomials, which is equivalent to the time-independent Shr\"odinger equation satisfied by the energy eigen-wavefunctions of the harmonic oscillator (see Appendix~\ref{app:hermite}):
\begin{equation}\label{eq:diffeq-Hermite}
   (-\partial_x^2 / 4 + x\partial_x) \chi_n(x) = n \chi_n(x) \,.
\end{equation}
Combining with \eqref{eq:QonHermite}, we obtain
\begin{equation}
      \mathbb{Q} | g(\{s_{a}\}) ) = \vert \mathrm{D} g  \rangle \,,\, \mathrm{D} = \sum_a 
       (-\partial_{s_a}^2 / 4 + s_a \partial_{s_a})  \,,
\end{equation}
for any function $g$. Equipped with this phase-space representation of the super-operator $\mathbb{Q}$, we are ready to evaluate the OTOC~\eqref{eq:Ct-Superop}. Using the definition of the inner product \eqref{eq:infiniteTinner} and the phase-space representation of the ensemble average~\eqref{eq:phasespacedef}, we have 
\begin{align} \label{eq:Ctphasespace1}
    \mathbf{C}(t) = C \int  
      f(t) \mathrm{D} f(t) \prod_a e^{- 2 s_a^2}  \mathrm{d} s_a \,,
\end{align}
where $C = (\pi / 2)^{-3/2}$, and $f(t) = f(\{s_a(t)\})$ with $s_a(t)$ being the time-evolved classical variables according to \eqref{eq:dsadt}. We can bring \eqref{eq:Ctphasespace1} to a more pleasant form by defining
\begin{equation}
    \Tilde{f}(t) := f(t) e^{-r^2}
\end{equation}
(recall $r^2 =\sum_a s_a^2$), in terms of which
\begin{align} \label{eq:Ctphasespace2}
    \mathbf{C}(t) = C \int  
      \tilde{f}(t) (-\Delta/4 + r^2) \tilde{f}(t)  \mathrm{d} s_a \,.
\end{align}
Now, the term proportional to $r^2$ is time-independent,  since both $r^2$ and the phase space volume are conserved by the classical dynamics~\eqref{eq:dsadt}. We can therefore write 
\begin{equation}\label{eq:OTOCphase-final}
    \mathbf{C}(t) = C_1  \int  
      | \nabla \tilde{f}(t)|^2   \mathrm{d} s_a + C_2 \,,
\end{equation}
where $C_1 = C / 4$ and $C_2 = C \int_{\mathbb{R}^3} \tilde{f}(0)^2 r^2 $ are both time-independent constants. Eq.~\eqref{eq:OTOCphase-final} is the advocated phase-space representation of the deep Hilbert space OTOC, and the main result of this section. It shows that the DHS OTOC measures the ${L}^2$ squared norm of the gradient of the time-evolved operator, represented as a function of phase space.
Thus, it is very similar to the phase space representation of the OTOC in the TSS. The latter, in the infinite-temperature ensemble of the TSS for example, is an integral on the two-sphere $ \mathbb{S}^2 = \{ \sum_a s_a^2 = 1\}$:
\begin{equation}
    \mathbf{C}_{\text{TSS}}(t) \propto \int_{\mathbb{S}^2}  | \nabla {f}(t)|^2 \,,\label{eq:OTOCphase-TSS}
\end{equation}
where $f(t)$ is the time-evolved operator $\mathcal{O}(t)$ represented as a function of phase space. 

It is now useful to contrast the DHS OTOC \eqref{eq:OTOCphase-final} with the TSS OTOC \eqref{eq:OTOCphase-TSS}. In both cases, the gradient's squared norm probes the sensibility of the classical trajectories to an initial perturbation. The only difference is that the phase space is $\mathbb{R}^3$ in the DHS case, and the two-sphere in the TSS case. This difference has significant consequences, as we shall see next.


\subsection{Super-exponential scrambling}\label{sec:super-exp-OTOC}
We now apply the phase-space representation \eqref{eq:OTOCphase-final} to show that the deep Hilbert space OTOC can grow super-exponentially. For the sake of concreteness, we will focus again on the LMG model~\eqref{eq:HLMG}. The scrambling of this model (as well as its kicked variants) has been well studied {in the TSS}. In particular, it is an example of ``saddle-dominated'' scrambling~\cite{scramblingvchaos}: the OTOC grows exponentially solely due to a saddle point in the phase space. 

The DHS OTOC growth is also saddle-dominated, because it has a similar phase space representation~\eqref{eq:OTOCphase-final}. More concretely, the classical dynamics~\eqref{eq:dsadt} given by the LMG Hamiltonian~\eqref{eq:HLMG}, $h = s_x + J s_z^2  / 2$, is such that the point $(s_x, s_y, s_z) = (r, 0, 0)$ is a fixed point. The linearized dynamics around it is 
\begin{equation}
    \dot{s_y} = (1 - J r) s_z + \dots \,,\, \dot{s_z} = -s_y \,.
\end{equation}
Therefore, for all $r > 1/J$, we have a saddle point, with the following instability exponent:
\begin{equation}
    \lambda_r = \sqrt{ J r - 1} \,.
\end{equation}
More precisely, $\dot{s}_+ = \lambda_r s_+ + \dots $ with $s_+ = \lambda_r^{-1} s_y + s_z$. Now we can adopt the ``saddle-dominated scrambling'' argument~\cite{scramblingvchaos} to estimate the DHS OTOC. The linearized dynamics $s_+(t) = e^{\lambda_r t} s_+$ is a good approximation of the true dynamics provided we start close enough to the saddle, i.e., in the  region 
$$ \mathcal{N} := \{ |s_+| < \delta e^{-\lambda_r t}, |s_-| < \delta, \sum s_a^2 = r^2 , r > 1/J \} \, $$
(where $s_-$ and $s_+$ form a local coordinate system of a neighborhood of $(r,0,0)$ of the sphere of radius $r$). Then the OTOC is dominated by this region of the phase space, and we have 
\begin{align}
    \mathbf{C}(t) &\sim  \int_\mathcal{N} |\nabla \tilde{f}(t)|^2 \nonumber \\
    &\sim   \int_{1/J}^\infty e^{-\lambda_r t} e^{2\lambda_r t} e^{- 2 r^2} \mathrm{d} r  \nonumber \\ 
    & \sim  \exp( c t^{a}) \,,\, a = \frac43  \quad \text{(LMG)}\label{eq:Ct-LMG}  \,.
\end{align}
In the second line, we wrote the three-dimensional integral as one over the radii. The factor $e^{-\lambda_r t}$ comes from the exponentially small width of $\mathcal{N}$ at radius $r$, and the $e^{2\lambda_r t}$ comes from the gradient squared applied to the fast changing function $s_+(t) = e^{\lambda_r t} s_+$; finally we performed a saddle point approximation of the $r$-integral ($c =  \frac38 J^{\frac23} $). This is the promised main result of this section: the deep Hilbert space OTOC grows super-exponentially.  As we can see from the above analysis, the parametrically fast dynamics at large $r$ is the origin of this anomaly, in a way similar to the explosive growth of the K-complexity. We should note that super-exponential OTOC growth has been reported in a kicked non-linear Schr\"odinger system, due to a different, non-equilibrium mechanism~\cite{superEG}.

 The exponent $4/3$ in \eqref{eq:Ct-LMG} is specific to the LMG model. Yet, the above method can be readily adapted to find the exponent of other models. For example, the Euler top with unequal couplings has saddle points~\footnote{The saddle points lie on the axis corresponding to the ``middle'' coupling constant; for example, if $J_x < J_y < J_z$, then $(0, \pm r, 0)$ is a saddle point for any $ r > 0$. In that case, $\omega = \sqrt{(J_y - J_x) (J_z - J_y)}$} with $\lambda_r = \omega r$, and therefore 
\begin{align}
    \mathbf{C}(t) \sim &
    \int_0^\infty e^{ \omega r t - 2 r^2}  
    \mathrm{d} r \nonumber \\ 
    \sim & 
    \exp(c t^a) \,,\, a = 2 \quad  \text{(Euler top)} \label{eq:Ct-Euler}
\end{align}

A few remarks are in order. The super-exponential OTOC growth does not require saddle-dominated scrambling (we considered such examples for simplicity). In fact, kicked/Floquet variants of LMG or Euler tops can display genuine classical chaos. In that case,  we expect the Lyapunov exponent to increase as a power law of $r$ ($r$ is still conserved by the classical dynamics), so the OTOC will grow super-exponentially as well if it is chaos-dominated. 

The super-exponential growth of OTOCs is consistent with the bound~\cite{hyp} relating K-complexity and OTOC growth: as we just saw, the K-complexity grows qualitatively faster (explosively) in all-to-all systems in the DHS. Thus, these systems are far from saturating the ``K-complexity $\gtrsim$ OTOC'' bound~\footnote{We recall from \cite{hyp} that the ``K-complexity $\gtrsim$ OTOC'' bound does not directly apply to the OTOC in the TSS~\eqref{eq:OTOCclassical}; one can only prove a relaxed version thereof. Meanwhile, the DHS OTOC does rigorously obey the usual bound, since it is a measure of operator size.}, unlike the SYK models (in the $q\to\infty$ limit). 

Finally let us point out that in a $k$-local all-to-all system with $k >2$, the analogue of the $r$-integral of \eqref{eq:Ct-Euler} may have an integrand $\sim e^{c r^{k-1} t - 2 r^2} $ that diverges at $r\to\infty$. This may result in a formally explosive OTOC growth which would only be regularized by finite $N$. We shall refrain from pursuing this possibility in the present work; as we shall see, the finite $N$ effect is already quite involved for $k=2$. 


\subsection{Finite $N$: pre-saturation and slow scrambling}\label{sec:finiteNOTOC}
\begin{figure}
    \centering
    \includegraphics[width=\columnwidth]{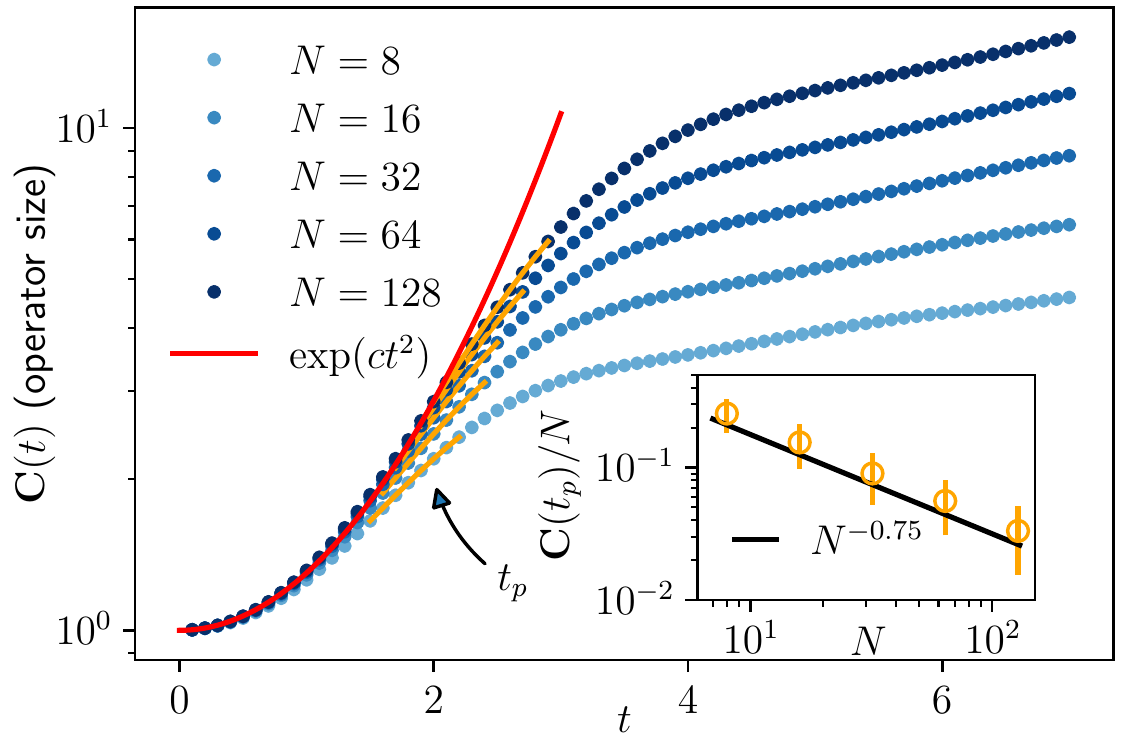}
    \caption{Numerical simulation of the DHS OTOC~\eqref{eq:OTOCintro} growth in the Euler top model $H = \sum_{jk} (J_xS_j^x S_k^x + J_y S_j^y S_k^y + J_z S_j^z S_k^z ) / (2\sqrt{N})$ with $ J_x = 0, J_y =- 1, J_z = 1/2$. The evolving operator is $\mathbf{O} = \mathbf{S}_x = N^{-1/2} \sum_j S_j^x$.  See Appendix~\ref{app:numerics} for methods. \textbf{Main}. The time-dependence of the OTOC for different system sizes $N$ is compared to the super-exponential growth prediction \eqref{eq:Ct-Euler} with $c = 0.26$ (fit). For each system size, a range for the estimated pre-saturation time scale $t_p$ is indicated by an orange interval; its extremities are the maximum of $\partial_t \ln C(t)$ and the minimum of $\partial^2_t \ln C(t)$, respectively. $C(t)$ becomes significantly $N$-dependent for $t \gtrsim t_p$, invalidating the semiclassical theory. \textbf{Inset}. The pre-saturation value  $C(t_p)$ (the error bar results from the width of the interval) as a function of $N$, compared to the power law $N^{-0.75}$, \eqref{eq:Ctp}. 
   } 
    \label{fig:dhs-otoc}
\end{figure}

In a finite system, the operator size cannot exceed the system size $N$, so the OTOC that measures it cannot grow indefinitely and must saturate. This saturation is usually characterized by a time scale $t_S$ (sometimes known as the scrambling time or Ehrenfest time), defined as the moment where the OTOC attains a finite fraction of its maximal value, $\mathbf{C}(t_S) \sim N$. We can however consider another time scale, $t_p$, the ``pre-saturation'' time, as when the finite-$N$ OTOC deviates significantly from the large $N$ limit (see caption of Fig.~\ref{fig:dhs-otoc} for a precise definition of $t_p$). The ``normal'' scenario of a finite size OTOC is one in which $t_p$ and $t_S$ coincide. The value of the OTOC at $t_p$ is a nonzero fraction of $N$ (as $N\to\infty$): 
\begin{equation}\label{eq:normalOTOC}
    \mathbf{C}(t_p) \sim N \quad \text{ (normal scenario)} \,.
\end{equation}
In other words, the finite-$N$ OTOC growth coincides with the $N\to\infty$ limit until it stops. The normal scenario is rather ubiquitous, and has been observed in SYK models as well as in semiclassical settings.

By contrast, the DHS OTOC in finite systems is \textit{anomalous}: it violates \eqref{eq:normalOTOC}. We show this by exact numerical calculation of $\mathbf{C}(t)$ in systems of various sizes (see Appendix~\ref{app:numerics} for methods). In both LMG and Euler top, it is visible that the finite-$N$ OTOC continues growing after it deviates from the $N\to\infty$ limit and becomes $N$-dependent. This qualitative observation already indicates that the normal scenario is not taking place in the DHS.

We now present two pieces of quantitative evidence to back up this statement. The first is a quantitative violation of \eqref{eq:normalOTOC} in the Euler top. In this model, the initial super-exponential growth regime $\mathbf{C}(t) \sim \exp(ct^2)$ is well established, so that we can identify the time scale $t_p$ and measure $\mathbf{C}(t_p)$ (see Caption of Figure~\ref{fig:dhs-otoc} for the practical method to do this). As a result, we find that 
\begin{equation}\label{eq:Ctp}
    \mathbf{C}(t_p) / N \sim N^{-b}, b\approx 0.75 \,,
\end{equation}
in contradiction with the normal scenario~\eqref{eq:normalOTOC}. 
As shown below, we obtained a result compatible with~\eqref{eq:Ctp} for the LMG model as well.

\begin{figure}
    \centering
    \centering
    \includegraphics[width=\columnwidth]{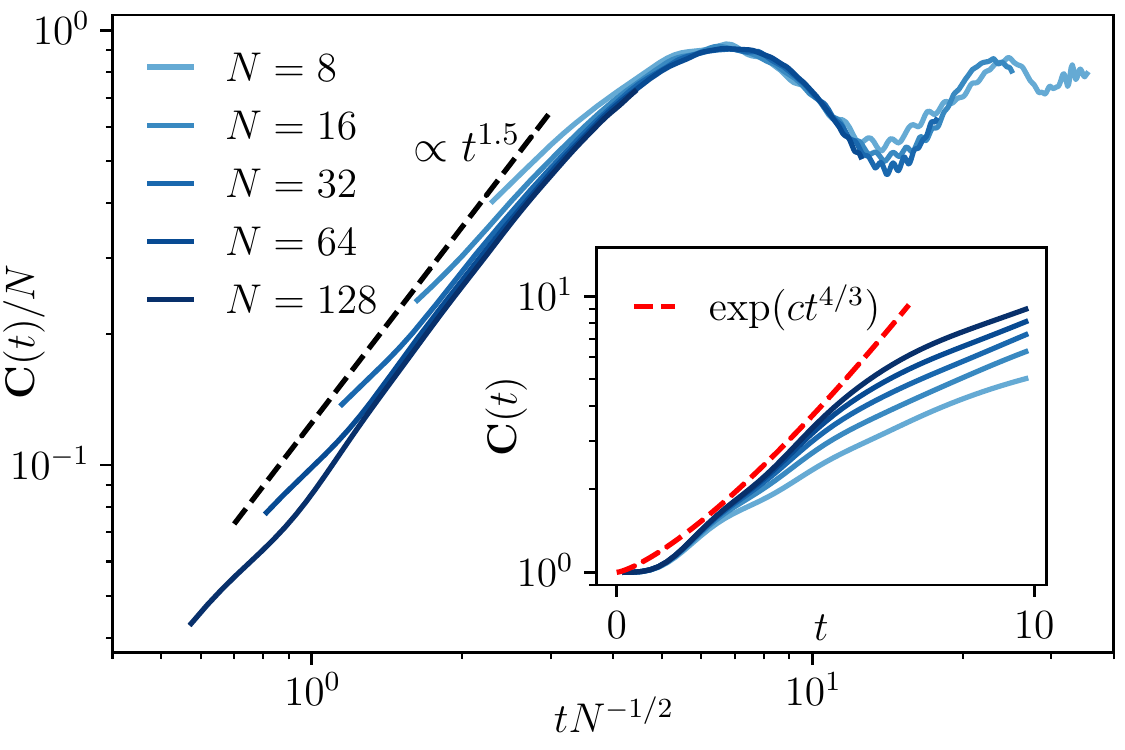}
    \caption{The DHS OTOC~\eqref{eq:OTOCintro} with $\mathbf{O} = \mathbf{S}_z = N^{-1/2} \sum_j S_j^z$ in the LMG model $H = \sum_{j} S_j^x + N^{-1/2} J \sum_{jk} S_j^z S_k^z / 2$ with $J = 1$. \textbf{Main}. The eventual saturation of the OTOC. We plot $\mathbf{C}(t)/N$ as a function of $t / N^{0.5}$, collapsing the data with different system sizes, except the initial growth regime. This confirms the saturation time scaling $t_S \sim N^{0.5}$~\eqref{eq:tS}. The intermediate-time growth is compared to the power law $t^{1.5}$, predicted by \eqref{eq:OTOCregimes} (with $b=0.75$). \textbf{Inset}. The same data plotted without rescaling, and restricted to $t \le 10$. The initial fast growth is consistent with the prediction \eqref{eq:Ct-LMG} with a fitted constant $c = 0.16$. }
    \label{fig:dhs-otoc-lmg}
\end{figure}

Next, we measure directly the scrambling time $t_S$ in the LMG model (with $\mathbf{O} = \mathbf{S}_z$), and find that it is surprisingly long. The long-time collapse in Figure~\ref{fig:dhs-otoc-lmg} indicates that
\begin{equation}
     t_S \sim N^{0.5}  \,. \label{eq:tS}
\end{equation}
We found the same scrambling time scaling for the Euler top~\footnote{Except for operators whose autocorrelation function decays slowly, such as $\mathbf{S}_z$ when $J_z$ is the minimal/maximal coupling. We observed numerically that these operators have a more involved, and slower, OTOC saturation.}. A power-law-in-$N$ scrambling time is incompatible with the normal scenario. Indeed, if the OTOC were to grow super-exponentially $\sim \exp(c t^a) $ until $\mathbf{C}(t) \sim N$ according to the normal scenario, the scrambling time would be $t_S \sim (\ln N)^{1/a}$, much shorter than we observed~\eqref{eq:tS}. 

 In summary, the numerical results unambiguously rule out the normal scenario for finite-system OTOC growth in the DHS. Instead, they indicate the following two-stage growth
 \begin{equation} \label{eq:OTOCregimes}
     \mathbf{C}(t) \sim \begin{cases}
         \exp(c t^a)  & t \lesssim t_p \sim ((1-b)/c \ln N)^{1/a} \\
         N^{1-b} t^{2b} &  t_p \lesssim t  \lesssim t_S \sim N^{1/2} \\
         N & t \gtrsim t_S \,.
     \end{cases}  
 \end{equation}
 Here we postulated (motivated by simplicity and numerical observation) that the intermediate time regime is described by a simple power law. Assuming this, one finds that the power of $t$ has to be $2b$ in order to match \eqref{eq:Ctp}, \eqref{eq:tS}, and $\mathbf{C}(t_S) \sim N$.
 Indeed, \eqref{eq:Ctp} implies that $t_p$ is proportional to $(\ln N)^{1/a}$, and grows more slowly than any power law of $N$. Thus, we may approximate it by $1$ when matching with the intermediate regime, which involves a much longer time scale (power-law in $N$). Namely, up to log corrections, $\mathbf{C}(t)$ grows from $N^{1-b}$ to $N$ as $t$ goes from $1$ to $N^{1/2}$. This fixes the exponent of $t$ to be $2b$. (The value of $b$ itself is not fixed by this argument.) In Figure~\ref{fig:dhs-otoc-lmg}, we compared the numerical data of the LMG model with a power law $t^{2b}$ corresponding to $b = 0.75$~\eqref{eq:Ct-LMG}, and find a reasonable agreement.

We speculate that the following mechanism could be behind the precocious end of the super-exponential growth regime. Recall that in that regime, the OTOC can be calculated as an integral over the phase space radial coordinate [see \eqref{eq:Ct-LMG} and \eqref{eq:Ct-Euler} above]:
 \begin{equation}
     \mathbf{C}(t) \sim \int e^{-2 r^2 + \lambda_r t} \mathrm{d} r \,,
 \end{equation} 
 This integral is dominated by the neighborhood of a saddle point $r_*$ that depends on $t$ but not on $N$. Now, here is the crucial heuristic input: the phase space spheres with radius $\sim r_*$ correspond to a quantum spin $s = \sqrt{N} r_* \ll N$, so we expect the OTOC contribution from $r\sim r_*$ to be $\ll N$ as well. Since the total OTOC is dominated by that region, this explains qualitatively why the $N\to\infty$ prediction fails when the OTOC is still $\ll N$. Quantitatively, our numerical data suggests that the $r\sim r_*$ contribution saturates at $N^{1-b} = N^{0.25}$~\eqref{eq:Ctp} in both Euler top and LMG models. To predict the value of the exponent would require a proper understanding of the quantization (finite-$N$ effect) of the DHS phase space. This is also necessary to  describe theoretically the intermediate time regime, and is left to future work.

To conclude our study on scrambling in the deep Hilbert space, let us emphasize the following: despite the super-exponential initial OTOC growth, all-to-all models in the deep Hilbert space are not super-fast scramblers. In fact they do not even qualify as ``regular'' fast scramblers, where by definition $t_S \sim \ln N$, despite their super-extensive many-body spectrum~\cite{yin-lucas}.

\section{Slow entanglement growth in the DHS}\label{sec:entanglement}
In this section, we consider the entanglement growth in a quantum quench from a product state in the DHS. This is a quite different quantity compared to those we studied so far, which are all essentially few point correlations in the infinite-temperature ensemble. So we will follow a different theoretical approach (in Sections \ref{sec:fieldtheory} through \ref{sec:effH}), which we preview here. Using the replica trick and a Hubbard-Stratonovich decoupling, we shall write the the $n$-th Renyi entanglement entropy as a path integral representation with an action proportional to $N$. We will contrast the case of an initial product state in the TSS with the case of an initial product state in the DHS, and it will become clear why the $1/N$ normalization \eqref{eq:HandSa} is the correct one for computing entanglement entropy in \textit{both} cases. We then evaluate the path integral using the Gaussian/semi-classical approximation, which is controlled by large $N$. This will be done by reverse engineering an effective free boson system that gives rise to the same path integral (up to quadratic approximation). For a TSS initial condition, the fictitious free boson system can be chosen to describe exactly the linearized dynamics along the classical phase-space trajectory, connecting our approach to the established ones in the literature~\cite{lerose-pappalardi2020pra,asplund16,bianchi18,hackl2018pra}. In particular, exponential instabilities give rise to linear-in-time entanglement entropy growth. On the other hand, DHS initial conditions typically lead to effective bosonic systems with only algebraic instabilities. As a result, we have a slow, logarithmic-in-time growth of entanglement entropy.

\subsection{Product state in the DHS}\label{sec:prodDHS}
Before studying entanglement, we shall characterize the initial state of our quench setup, which is a product state:
\begin{equation}\label{eq:defPsi0}
    \vert \Psi_0 \rangle = \prod_{j=1}^N \vert s_j \rangle \,.
\end{equation}
Here, $\vert s_j \rangle$ is a spin-$1/2$ corresponding to the pointer $s_j \in \mathbb{S}^2$ on the Bloch sphere, such that 
\begin{equation}\label{eq:sj-def}
   \langle s\vert 2 S^a \vert s \rangle = s_a  \,.
\end{equation}   
($\vert s_j \rangle$ is defined up to a phase, bur the phase ambiguity will not affect the entanglement entropy). To have a well-defined $N\to\infty$ limit, we require that the set of pointers $\{s_j\}_{j=1}^N$ tends to a limiting distribution $P(s)$ on the Bloch sphere:
\begin{equation}
    \frac1N \sum_{j=1}^N \delta(s - s_j) \longrightarrow P(s)  \,. \label{eq:defPn}
\end{equation}  
We shall consider two types of initial conditions. The first is those in the TSS, for which all $s_j \equiv s$ are equal; in that case the distribution is a delta peak. This case has been well studied in the literature and we will cover it as a ``control group''. 

The second type is those in the DHS, for which $P(s)$ is a smooth distribution, i.e., not a discrete set of delta peaks, whose centre of mass is at the origin:
\begin{equation}
    \sum_{j=1}^N s_j = 0   \quad \text{(DHS)} \,.  \label{eq:njsum}
\end{equation}
Examples include the uniform distribution on the Bloch sphere, and that on a great circle thereof. 
We should note that this product state is somewhat atypical within the DHS since a typical state in the DHS would be Haar-random and thus have volume law entanglement (with respect to any bipartition). However, our product state does indeed lie in the DHS according to our earlier definition based on the expectation value of collective variables \eqref{eq:defDHS}, as we now show:

\noindent \textbf{Proposition}. With respect to the state \eqref{eq:defPsi0} under the condition \eqref{eq:njsum}, the DHS collective spins $\mathbf{S}_a$ behave as Gaussian variables with vanishing mean and the following covariance: 
\begin{align}
    &\langle \Psi_0 \vert  \mathbf{S}_a  \mathbf{S}_b    \vert \Psi_0 \rangle \stackrel{N\to\infty} =  \frac14 (\delta_{ab} - \overline{s_a s_b}^s) \,, \label{eq:prodDHS-res}
\end{align}
where here and below, 
\begin{equation}
    \overline{f(s)}^s := \int_{\mathbb{S}^2} P(s) f(s) \mathrm{d}^2 s
\end{equation}
denotes an average over the distribution $P(s)$.

The above proposition is proved in Appendix~\ref{app:traces}. To interpret it, recall that $\mathbf{S}_a$ behave also as Gaussian random variables in the infinite-temperature $\rho_{\infty}$ of the DHS. Thus, \eqref{eq:prodDHS-res} shows that the quantum fluctuations of $\mathbf{S}_a$ in the DHS product state are of the same order of magnitude (smaller by a factor of order 1) as their quantum-statistical fluctuation in $\rho_{\infty}$. By contrast, if a product state defined by \eqref{eq:defPsi0} violates the condition \eqref{eq:njsum}, some of the DHS collective variables would acquire large expectation values $\sim \sqrt{N}$, while the TSS ones $\mathcal{S}_a$ have order one expectation values. In this sense,  the product state  $ \vert \Psi_0 \rangle  \langle \Psi_0 \vert $ satisfying the condition \eqref{eq:njsum} is more similar to the DHS ensemble $\rho_\infty$ than to a TSS state, and it is reasonable to call it a ``DHS product state''. In particular, an OTOC evaluated on the state $ \vert \Psi_0 \rangle  \langle \Psi_0 \vert $ (in lieu of $\rho_{\infty}$) will also grow super-exponentially until pre-saturation, by the same argument of the previous section. 


%

\subsection{Path integral for entanglement}\label{sec:fieldtheory}
We consider the bipartite entanglement of the time-evolved state $\vert \Psi_t \rangle = e^{-i H t} \vert \Psi_0 \rangle$. More concretely, we split the spins-$1/2$'s into two groups $\{1, \dots, N\} = A \cup B$ of comparable size: $|A| = x N, |B| = (1 - x) N$, with $x \in (0,1)$ fixed as $N\to\infty$. For simplicity, we shall assume that the distribution $\{s_j, j \in A\}$ and $\{s_j, j \in B\}$ tend both to $P(s)$ as $N \to \infty$; one could think of the bi-partition as being randomly chosen, independently of $s_j$. Recall that the $n$-th Renyi entropy $S_n$ is defined as~\footnote{The index $n$ is reserved for the Renyi index and replica number, so $S_n$ should not be confused with a spin.}:
\begin{equation}\label{eq:Renyidef}
   S_n :=  \frac{1}{1-n} \ln  \left( \mathrm{Tr}[\rho_A^n ] \right) \,,
\end{equation}
where $\rho_A$ is the reduced density of the subsystem $A$, $\rho_A = \mathrm{Tr}_B \vert \Psi_t \rangle \langle \Psi_t \vert$. The von Neumann entanglement entropy is given by the $n \to 1$ limit of the Renyi one. 

In the rest of this section, we derive an exact path integral representation of the Renyi entropy for $n = 1, 2, 3, \dots$. For concreteness, we shall focus on the LMG model, although our method applies to any Hamiltonian that is at most quadratic in the collective variables $\mathcal{S}_a$, e.g. the Euler Top (see Sec.~\ref{sec:euler-ent}). The basic idea is to apply the Hubbard-Stratonovich decoupling to the infinitesimal time evolution operator 
\begin{align}
   e^{\mp i  H   \mathrm{d} t} &= e^{ \mp i  S  (\mathcal{S}_x + J \mathcal{S}_z^2 / 2)  \mathrm{d} t } \nonumber \\ 
    &=  \int [\mathcal{D} \varphi]  e^{\mp i S  (-\frac{1}{2 J} \varphi^2 + \mathcal{S}_z \varphi + \mathcal{S}_x )  \mathrm{d} t} \nonumber \\
    &= \int [\mathcal{D} \varphi]  e^{\pm i S  \frac{1}{2 J} \varphi^2 \mathrm{d} t} \prod_j e^{\mp i (S_j^z \varphi + S_j^x) \mathrm{d} t} \label{eq:decouple1}
\end{align}
Here, the integral measure is $[\mathcal{D} \varphi] =  (\pm i 2\pi J / (S \mathrm{d} t))^{-1/2} \mathrm{d} \varphi $ and the integral contour of $\varphi$'s is suitably chosen so that the Gaussian integral converges. In the third line, we recall that $\mathcal{S}_a = \sum_j S_j^a/S$. As a result, we get a factorized operator for fixed $\varphi$. Applying the same decoupling to all the infinitesimal time evolution factors involved in the density matrix at $t$, we obtain 
\begin{align}
    \rho = \vert \Psi_t \rangle \langle  \Psi_t \vert = &\int   [\mathcal{D} \varphi_{\pm}(t)]  e^{i S  \int_0^t  \frac{1}{2 J} (\varphi_{+}^2 - \varphi_{-}^2) \mathrm{d} t} \nonumber \\  & \times  \prod_{j=1}^N \left( U_{\varphi_+}(t,0) \vert s_j \rangle \langle s_j \vert  U_{\varphi_-}(0,t) \right)
\end{align}
where 
\begin{align} \label{eq:Uvarphi}
   U_{\varphi}(u,v)  = \mathcal{T} e^{ - i \int_u^v H_{\varphi}(w) \mathrm{d} w }
\end{align}
is the time-evolution operator on a single qubit, under a time-dependent Hamiltonian
\begin{equation} \label{eq:Hvarphi}
     H_{\varphi}(w) := \varphi(w) S^z  + S^x \,. 
\end{equation}
controlled by the field $\varphi$. Note that we have introduced $\varphi_{+}$ and $\varphi_{-}$ for the evolution of the ket and bra, respectively, which is common practice in (non-equilibrium) Keldysh field theory~\cite{kamenev_2011}.

\begin{figure}
    \centering
    \includegraphics[width=1\columnwidth]{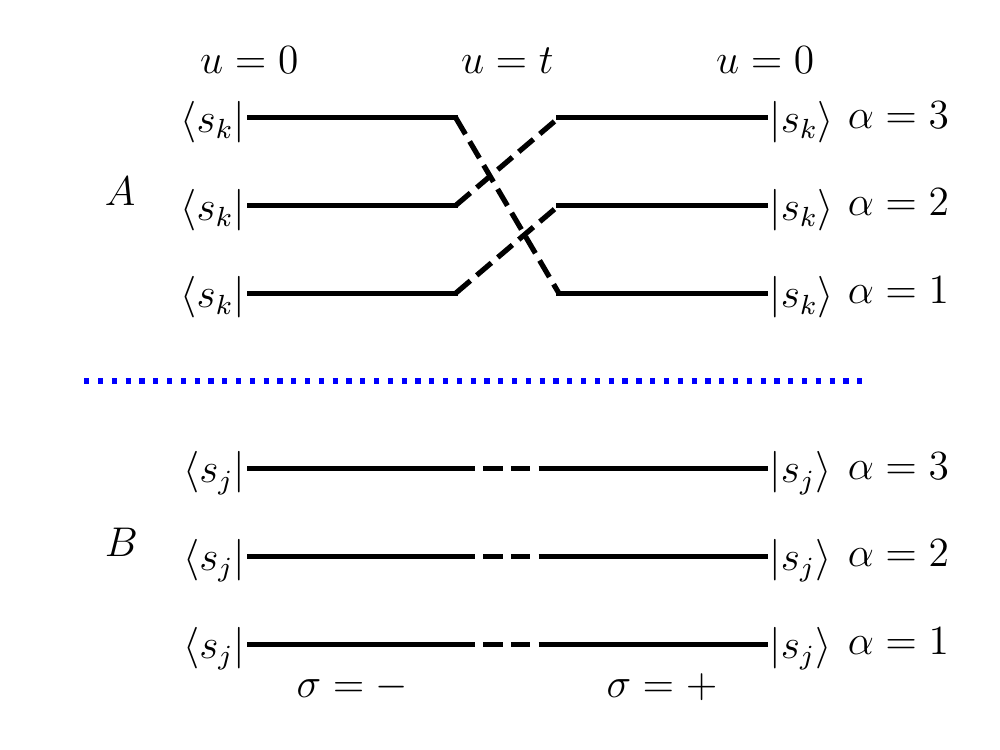}
    \caption{Illustration of the path integral contour with replica number $n=3$. The forward contours ($\sigma = +$, the time $u$ increases from $0$ to $t$) start from the right, and are connected to the backward contours ($\sigma = -$, $u$ decreases from $t$ to $0$) which end at the left. In the subsystem $A$ (and not $B$), a cyclic permutation is applied when connecting the forward and backward contours, as indicated by the dashed lines.  }
    \label{fig:keldysh}
\end{figure}

To compute the Renyi entropy, we need $n$ replicas of $\rho$, and contract the ket at site $j$ of the $\alpha$-th replica with the bra of the same replica if $j \in B$, and of the $(\alpha +1)$-th replica otherwise ($n + 1 \equiv 1$). See Figure~\ref{fig:keldysh} for an illustration. This will give rise to a path integral over replicated fields $\varphi_{\alpha  \pm}$, as follows:
\begin{align}
    \mathrm{Tr}[\rho_A^n] =& 
    \int [\mathcal{D} \varphi_{\alpha\pm}(t)] 
    e^{i S  \int_0^t \sum_{\alpha} \frac{1}{2 J} (\varphi_{\alpha+}^2 - \varphi_{\alpha-}^2) \mathrm{d} t} \nonumber \\  & 
    \times  \prod_{j \in B}  \prod_{\alpha=1}^n F[\varphi_{\alpha +}, \varphi_{\alpha -} ,s_j] \nonumber \\
    &
  \times   \prod_{j \in A }  \prod_{\alpha=1}^n F[\varphi_{\alpha +}, \varphi_{(\alpha+1) -} ,s_j] 
\end{align}
where
\begin{equation} \label{eq:defF}
    F[\varphi_+, \varphi_- , s] := \langle s \vert U_{\varphi_-}(0,t) U_{\varphi_+}(t,0) \vert s \rangle \,.
\end{equation}
Now, in the large $N$ limit, we can turn the products into the exponential of $N = 2S$ times an average over the distribution $P(s)$. Thus, we finally obtain a path integral representation of $ \mathrm{Tr}[\rho_A^n] $ with a large $S$ action:
\begin{equation} \label{eq:pathintegral}
    \mathrm{Tr}[\rho_A^n] = 
   \int [\mathcal{D} \varphi] e^{i S \mathcal{A}[\varphi] } \,, \,  \mathcal{A}=  \mathcal{A}_0 +  \mathcal{A}_1 \,,\,
\end{equation}
where 
\begin{align} \label{eq:action-renyi}
\mathcal{A}_0[\varphi] =&\frac1{2J}  \int_0^t \mathrm{d} u  \sum_{\alpha=1}^n
\left[ \varphi_{\alpha +}^2(u) - \varphi_{\alpha -}^2(u) \right]  \,,\, \\ \mathcal{A}_1[\varphi] =&  - 2 i    (1-x) \overline{  \ln F[  
\varphi_{\alpha +}, \varphi_{\alpha -}, s ] }^s  \nonumber \\
&  - 2 i x  \, \overline{  \ln F[ 
\varphi_{\alpha +}, \varphi_{ (\alpha+1) -}, s ]  }^s \,. \nonumber 
\end{align}
where we recall that $x = |A|/N$ is the relative size of the subsystem $A$.

Eq. \eqref{eq:action-renyi} is derived for the LMG model. However, the method can be adapted to a general Hamiltonian that is quadratic in $\mathcal{S}_a$: it suffices to make the field $\varphi$ multi-component in order to decouple the quadratic form. For instance, for the Euler top (see Section~\ref{sec:euler-ent} below), the path integral will be over the field $\varphi_{\alpha\pm}^a$, $a = x, y,z$. The first term of the action \eqref{eq:action-renyi} will sum over $a$ with $J \to J_a$, and \eqref{eq:Hvarphi} will become $\sum_a \varphi^a S^a$. A general quadratic form in $\mathcal{S}_a$ can be diagonalized and then treated in the same way. In what follows, we will focus on the LMG case \eqref{eq:action-renyi}. 

Before proceeding, we remark that using the $1/N$ normalization \eqref{eq:HandSa} is crucial to obtain a large $S$ action; had we used the $1/\sqrt{N}$ one \eqref{eq:HDHS}, the action would have had a part proportional to $S$ and the other to $\sqrt{N}$. This is the formal way to see that the $1/N$ normalization is the correct one to study the entanglement growth from a product state, both in and away from the TSS. 

\subsection{Semiclassical analysis}\label{eq:semiclassical}
We now proceed to a semiclassical analysis of the path integral above, i.e., we approximate the latter as a Gaussian integral over a particular saddle point of the action (in field-theory jargon we evaluate the path integral up to one-loop). The semiclassical expansion is controlled in the large $S$ limit, for fixed $t$. Thus, our analysis aims to capture the time regime where the entanglement growth has not yet saturated due to finite $S$ (we will numerically study the saturation, see below). 

\subsubsection{Classical equation of motion}
We start by looking at the classical equation of motion of \eqref{eq:action-renyi}. We will analyze this under an assumption (to be justified below; see \eqref{eq:action-vanish}) --- namely, we will evaluate the functional derivative on configurations with equal components:
\begin{equation}\label{eq:condition}
     \varphi_{\alpha+} = \varphi_{\alpha-} =: \varphi^{\text{cl}}\,.
\end{equation} 
A consequence of~\eqref{eq:condition} is that functional derivatives of $F$ \eqref{eq:defF} are essentially equal to correlation functions on a Keldysh contour. In particular, one can check that:
\begin{equation}
   \left. \frac{\delta \ln F[\varphi_+, \varphi_-, s]}{\delta \varphi_\pm(u)} \right|_{\varphi_\pm =\varphi^{\text{cl}}} = \mp i \left< S^z(u) \right>_s
\end{equation}
where
\begin{align}
  \left< \dots \right>_s :=  \langle s \vert \dots \vert s \rangle  \,,\, S^z(u)  :=  U_\varphi (0,u)  S^z U_\varphi (u,0)   \,.
\end{align}
That is, the first functional derivative is the expectation value (one-point function) of $S^z$, under the evolution of $H_\varphi$. Therefore, by \eqref{eq:action-renyi}, the classical equation of motion reads:
\begin{equation} \label{eq:EOM}
    \varphi^{\text{cl}}(u) / J = \overline{\left< 2 S^z(u) \right>_s}^s \,.
\end{equation}
This equation has a simple interpretation in terms of ``mean-field'' classical spin dynamics. Since $H_{\varphi}$ is linear in the spins, the evolution of the spin-$1/2$ is given by the following classical dynamics of its pointer $s$ on the Bloch sphere:
\begin{equation}\label{eq:classical-ent}
    \partial_u {s}_a(t) = \{ s_x + s_z \varphi^{\text{cl}}, s_a (u)\}_{\text{P.B.}} \,.
\end{equation}
Then \eqref{eq:EOM} identifies the classical solution to the average $z$-component of the time-evolved pointer distribution: 
\begin{equation} \label{eq:phiclissz}
    \varphi^{\text{cl}}(t) / J = \overline{ s_z(t)}^s \,.
\end{equation}

For an initial state in the TSS, the distribution is concentrated on a simple pointer $s$, which evolves under \eqref{eq:classical-ent}. Combined with \eqref{eq:phiclissz}, we  have
\begin{align}
     \dot{s}_a(t) &= \{ s_x + s_z \varphi^{\text{cl}}, s_a (t)\}_{\text{P.B.}} \nonumber \\ &= \{ s_x + J s_z^2  /2  , s_a (t)\}_{\text{P.B.}} \,,
\end{align}
that is, the pointer evolves under the classical dynamics~\eqref{eq:dsadt} given by the LMG Hamiltonian. In particular, if the initial condition $s = (1,0,0)$ is the fixed point of the LMG classical dynamics, the solution has simply 
\begin{equation}\label{eq:phi-cl-TSS}
    \varphi^{\text{cl}}(t) \equiv 0 \,,\, \text{TSS, } s = (1,0,0) \,.
\end{equation}
This case will be of interest since the dynamical instability around the fixed point leads to a linear growth in entanglement entropy, see below.

For an initial state in the DHS satisfying \eqref{eq:njsum}, \eqref{eq:classical-ent} and \eqref{eq:phiclissz} are solved by 
\begin{equation}
    \varphi^{\text{cl}}(t) = 0   \quad \text{(DHS)}  \,.
\end{equation}
Indeed, this implies that classical Hamiltonian is $s_x$, which rotates the distribution of the pointers, and keeps the center of mass at zero.

In all cases, it is straightforward to check that the action~\eqref{eq:action-renyi} vanishes at the above classical saddle point:
\begin{equation}\label{eq:action-vanish}
    \mathcal{A}[\varphi^{\text{cl}}] = 0 \,.
\end{equation}
Indeed, a sufficient condition for this is the assumption~\eqref{eq:condition}, which guarantees that $\ln F = 0$ and the forward and backward contribution to $\mathcal{A}_0$ cancel each other. Therefore, to compute the Renyi entropy, we shall integrate over fluctuations around the saddle point (this is the subject of the next section.) The vanishing of the saddle action ensures that the resulting Renyi entropy will approach $0$ as $t \to 0$, which it should be. It is thus highly unlikely for other saddle points to contribute, except for an amount that is exponentially small in $N$. This justifies our assumption \eqref{eq:condition}: the classical saddle $\varphi^{\text{cl}}$ with identical values on all Keldysh folds gives the dominant contribution to $\mathrm{Tr}[\rho_A^n]$ in the large $N$ limit for fixed $t$. Physically, this amounts to saying that the entanglement entropy growth before its finite $N$ saturation is given by quantum fluctuations around the classical dynamics, which, as we see above, is captured by $\varphi^{\text{cl}}$. 

\subsubsection{One-loop determinant}\label{sec:onloop}
We now evaluate the path integral by approximating the action up to quadratic order in $\varphi - \varphi^{\text{cl}}$. This gives us a ratio of determinants:
\begin{equation}
 \int [\mathcal{D} \varphi] e^{i S \mathcal{A}[\varphi]} \approx 
  \frac{\det(\mathcal{H}_0)^{\frac12}}{\det(\mathcal{H})^{\frac12}}  \,. \label{eq:ratiodet}
\end{equation}
Here, $\mathcal{H}$ is the Hessian of the action \eqref{eq:action-renyi}:
\begin{equation}
    \mathcal{H} = \frac{\delta^2 \mathcal{A}}{\delta \varphi_{\alpha_1 \sigma_1}(t_1) \delta \varphi_{\alpha_2 \sigma_2}(t_2)} \,.
\end{equation}
It is a matrix with indices $\alpha_{1,2} = 1,\dots,n$, $\sigma_{1,2} \in \{ +, -\}$ and $t_{1,2} \in [0,t]$ (to simplify notation, we will suppress these indices in $\mathcal{H}, \mathcal{H}_0$ and $\mathcal{K}$ below).  $\mathcal{H}^0$ is the Hessian of $\mathcal{A}_0$, and is equal to:
\begin{align} \label{eq:H0}
   \mathcal{H}_0 = 
   \frac{\sigma_1}J \delta(t_1 - t_2) \delta_{\alpha_1, \alpha_2}\delta_{\sigma_1, \sigma_2} \,.
\end{align}
This is because the integration measure was chosen such that the path integral of the quadratic part is one, see \eqref{eq:decouple1} above. 

To further evaluate \eqref{eq:ratiodet}, we write 
\begin{equation}
    \frac{\det(\mathcal{H}^0)}{\det(\mathcal{H})}  = 1/\det(\mathcal{H} \mathcal{H}_0^{-1}) \,.
\end{equation} 
$\mathcal{H}_0$ is diagonal \eqref{eq:H0}, and simple to invert. Combined with the definition of the Renyi entropy, we find
\begin{equation}\label{eq:Renyidet}
(n-1) S_n =  \frac12 \ln \det \left( I - i J \mathcal{K} \right) \,,
\end{equation}
where 
\begin{align} 
& \mathcal{K} =
    ((1-x) \delta_{\alpha_1,\alpha_2} +  x 
   \delta_{\alpha_1+1,\alpha_2}) G_{\sigma_1\sigma_2}(t_1, t_2)  \sigma_2 \,, \label{eq:KandG} \\
& G_{\sigma_1 \sigma_2}(t_1, t_2) := 2  \overline{\left.\frac{\delta^2 \ln F[\varphi_+, \varphi_-, s]}{\delta \varphi_{\sigma_1}(t_1) \delta \varphi_{\sigma_2}(t_2) }\right\vert_{\varphi_\pm = \varphi^{\text{cl}}}}^s \,. \label{eq:Gdef}
\end{align}
Here $\mathcal{K}$ is the nontrivial part of the determinant (resulting from the entangling interaction), and is essentially built from the Green function $G$.
Indeed, the latter is another functional derivative of $F$, whose calculation is simplified when evaluated on configurations with $\varphi_+ = \varphi_-$. We obtain the connected time-ordered two-point correlator (averaged over $s$),
\begin{align}\label{eq:Greenfunc}
  &   G_{\sigma_1 \sigma_2}(t_1, t_2)  \\ 
  = &
     2 \left(\overline{ \left< \mathcal{T}_{\sigma_1\sigma_2} S^z(t_1) S^z(t_2) \right>_s - \left< S^z(t_1) \right>_s \left< S^z(t_2) \right>_s}^s \right)  \nonumber
\end{align}
where the time ordering $\mathcal{T}_{\sigma_1\sigma_2}$ is done on the Keldysh contour, that is, 
\begin{subequations}
\begin{align}
  &  \mathcal{T}_{+-} A(t_1) B(t_2) = B(t_2)  A(t_1)  \\
  &  \mathcal{T}_{-+} A(t_1) B(t_2) =  A(t_1) B(t_2)   \\
  &    \mathcal{T}_{++} A(t_1) B(t_2) = \begin{cases}
       A(t_1) B(t_2) &  t_1 > t_2 \\
      B(t_2) A(t_1) &  t_2 > t_1
  \end{cases} \\
   &    \mathcal{T}_{--} A(t_1) B(t_2) =  \begin{cases}
       A(t_1) B(t_2) &  t_1 < t_2 \\
      B(t_2) A(t_1) &  t_2 < t_1
  \end{cases}  
\end{align}
\end{subequations}
These correlation functions depend on the classical configuration $\varphi^{\text{cl}}$. In general they do not have a simple expressions. In what follows, we shall focus on a few instances where $ \varphi^{\text{cl}} \equiv 0$, such that explicit calculation can be done simply. In all cases, we have $H_{\varphi} = S^x$, so 
\begin{equation}
    S^z(u) = S^z \cos(u) + S^y \sin(u) \,. 
\end{equation} 
Then it is straightforward to compute \eqref{eq:Greenfunc} by applying the time-ordering rules and averaging over $s$. As a result, we find 
  \begin{subequations} \label{eq:Ggen}
    \begin{align}
        &2G_{++}  = c_{12} + i |s_{12}|  \overline{x} - c_1c_2  \overline{z^2} - s_1 s_2  \overline{y^2} - s_{12} \overline{y z} \\
        &2G_{--} = c_{12} - i | s_{12}|  \overline{x} - c_1c_2  \overline{z^2} - s_1 s_2  \overline{y^2} - s_{12} \overline{y z}\\
        &2 G_{-+} = c_{12} + i s_{12}  \overline{x} - c_1c_2  \overline{z^2} - s_1 s_2  \overline{y^2} - s_{12} \overline{y z}   \\
        & 2G_{+-} =  c_{12} - i s_{12}  \overline{x} - c_1c_2  \overline{z^2} - s_1 s_2  \overline{y^2} - s_{12} \overline{y z}
    \end{align}
    \end{subequations}
where we used the shorthand notations $
 c_{12} \equiv \cos(t_1 - t_2)$, $s_{12} \equiv \sin(t_1 - t_2)$, $c_j = \cos t_j$, $ s_j = \cos t_j$, $x,y,z = s^{x}, s^{y}, s^z$, $\overline{[\dots]} = \overline{[\dots]}^s$. 
These results hold for any distribution $P(s)$. Let us specify a few examples that we will focus on in what follows:
\begin{enumerate}
    \item For the TSS initial state with the pointer located at the fixed point $s_j \equiv (1,0,0)$ of the LMG classical dynamics, we have  $
        \overline{x} = 1 \,,\, \overline{z^2} = \overline{y^2} = \overline{yz} = 0 \,,$
    and therefore 
    \begin{equation}\label{eq:G-TSS}
        G_{++} = {G_{--}^*}  = \frac{e^{i|t_{12}|}}2 \,,\, G_{-+} = G_{+-}^* = \frac{e^{i t_{12}}}2 
    \end{equation}
    where $*$ denotes the complex conjugate. 
    \item When $s$ is uniformly distributed on the Bloch sphere, we have $
          \overline{x} = 0 \,,\, \overline{z^2} = \overline{y^2} = \frac13 \,,\,   \overline{yz} = 0 \,. $ So
     \begin{equation}
         G_{\sigma_1 \sigma_2}(t_1, t_2) = \frac{1}{3} \cos(t_{12}) \,, \label{eq:Gcase2}
     \end{equation}
     for any $\sigma_1, \sigma_2$. 
     \item Another simple example (that is more convenient in finite size numerics) is one where  $s$ is uniformly distributed on the great circle with $x = 0$. This is similar to the previous example, except that $ \overline{z^2} = \overline{y^2} = \frac12$. Thus
     \begin{equation}
         G_{\sigma_1 \sigma_2}(t_1, t_2) = \frac{1}{4} \cos(t_{12}) \,,\label{eq:Gcase3}
     \end{equation}
     for any $\sigma_1, \sigma_2$. 
\end{enumerate}

\subsection{Effective Hamiltonian}\label{sec:effH}
The determinant expression~\eqref{eq:Renyidet} has an obvious drawback: we cannot read off the qualitative entanglement growth behavior --- for example, whether the growth is linear or logarithmic in $t$ --- directly from the kernel $\mathcal{K}$.  


In this section, we shall do this analytically by solving a ``reverse-engineering'' problem. That is, we find a quench setup in a few-body bosonic system with a quadratic Hamiltonian (which we call the \textit{effective Hamiltonian}) and a Gaussian initial state. We shall apply a similar field-theoretical treatment as above to the bosonic Hamiltonian to calculate the Renyi entanglement entropy of the boson setup in terms of a determinant, and show that its bipartite entanglement entropy is exactly given by the RHS of \eqref{eq:Renyidet}, as long as the effective Hamiltonian is appropriately chosen.

By finding an effective bosonic Hamiltonian, we reduce the problem to the solved one of calculating entanglement in a free boson model~\cite{bianchi18,lerose-pappalardi2020pra}. This is a known instance where a Zurek-Paz type relation holds~\footnote{We stress that the reduction of the DHS entanglement calculation to a free boson model does not imply a Zurek-Paz relation in the DHS. Because the DHS OTOC growth is not related to the same free boson model and its Lyapunov exponents.}. Namely, the asymptotic entanglement growth behavior can be obtained simply from the stability of the linear dynamics, which is described by a dynamical matrix. When the latter has eigenvalues (local Lyapunov exponents) with positive real part, the entanglement grows linearly, with a rate given by the sum of positive Lyapunov exponents. When the dynamical matrix is not diagonalizable and has Jordan blocks (of size larger than 1), there will be logarithmic corrections~\cite{lerose-pappalardi2020pra,pappalardilerosePRR}. These turn out to dominate the the entanglement growth from the DHS initial states that we shall consider. 

In what follows, we shall first illustrate the method for example 1 (TSS at a fixed point), as a benchmark. Then we apply it to the DHS examples 2 and 3, for which the results are new. 

\subsubsection{TSS (warm-up) example}
In the case of the TSS state corresponding to $s = (1,0,0)$, the effective Hamiltonian can be guessed by the Holstein–Primakoff transformation, together with some consideration to account for the bi-partition. As a result, we propose the following effective Hamiltonian acting on two degrees of freedom:
\begin{align}\label{eq:Heff-TSS}
 &H =  H_{0A}  + H_{0B} + \frac{J}2 (\sqrt{x} p_A + \sqrt{1-x} p_B)^2  \,,\,  \\
 & H_{0A} = - \frac12 ( q_A^2 + p_A^2) \,,\, 
 H_{0B} =  - \frac12 ( q_B^2 + p_B^2) \,.  \label{eq:H0-TSS}
\end{align}
 Here, $q_{A}, p_A$ and $q_B, p_B$ are two independent canonical position-momemtum pairs:
\begin{equation}
    [q_A, p_A] = i  \,,\, [q_B, p_B] = i  \,.
\end{equation}
We also recall that $x \in (0,1)$ is the relative size of the subsystem $A$ in the original quench setup. In our effective problem, the initial state will be the ground state of $-H_{0A} - H_{0B}$. We will consider the evolution of the Renyi entanglement entropy with respect to the bipartition $A \sqcup B$; note that the initial state is factorized:
\begin{equation}\label{eq:initialstate}
    \vert \Psi_0  \rangle = \vert 0 \rangle_A \vert 0\rangle_B \,.
\end{equation}

We argue that the Renyi entropy is exactly given by the determinant \eqref{eq:Renyidet}, supplemented with \eqref{eq:KandG}, \eqref{eq:Ggen} and \eqref{eq:G-TSS}.  For this, we apply the Hubbard-Stratonovich transform to the term proportional to $J$ in \eqref{eq:Heff-TSS}.  Following almost the same steps as before, we may obtain the same path integral representation \eqref{eq:pathintegral}, except that the interacting action becomes
\begin{align}
 &   \mathcal{A}_1[\varphi]=- i   \sum_{\alpha=1}^n \left(  \ln F_A[ 
\varphi_{\alpha +}, \varphi_{\alpha -}]  +  \ln F_{B}[ 
\varphi_{\alpha +}, \varphi_{ (\alpha+1) -}]  \right)  \,, \nonumber \\
 & F_A[ \varphi_{+}, \varphi_{-}] = \langle 0_A \vert 
 U^A_{\varphi_-}(0,t) U^A_{\varphi_+}(t,0) 
 \vert 0_A \rangle \,,\, \nonumber  \\ 
 & U^A_\varphi(v,u) = \mathcal{T} e^{-i \int_u^v H^A_{\varphi}(w) \mathrm{d} w } \,,  \nonumber \\ & H^A_{\varphi} = - \frac12 ( q_A^2 + p_A^2) + \sqrt{x} p_A \varphi \,,
\end{align}
and similarly with $A \to B$, $x \to 1-x$. Then, we check that $\varphi^{\text{cl}} \equiv 0$ is a classical saddle point of the total action [this matches \eqref{eq:phi-cl-TSS} above], and evaluate the path integral by a semiclassical (Gaussian) approximation around it. Since we are dealing with free bosons, the approximation is exact. It is not hard to check that \eqref{eq:Renyidet}, \eqref{eq:KandG} are still correct if we change \eqref{eq:Greenfunc} to 
\begin{equation}
     G_{\sigma_1 \sigma_2}(t_1, t_2)  
  =  \left<  0 \vert \mathcal{T}_{\sigma_1\sigma_2} p(t_1) p(t_2) \vert 0 \right>  \,,
\end{equation}
where $p(u) = p \cos(u) + p \sin(u) $ evolves under $H_{0} =  - \frac12 ( q^2 + p^2) $ (since $\varphi = \varphi^{\text{cl}} = 0$). An elementary harmonic oscillator calculation shows that the Green functions exactly coincide with \eqref{eq:G-TSS}. This concludes the demonstration that the effective Hamiltonian \eqref{eq:Heff-TSS} and the initial state~\eqref{eq:initialstate} is semiclassically equivalent to the TSS setup (case 1 in Section~\ref{sec:onloop}). 

Equipped with this equivalence, we can readily understand the entanglement growth using known results~\cite{asplund16,bianchi18,hackl2018pra,lerose-pappalardi2020pra}, which states that the entanglement entropy grows linearly if the quadratic Hamiltonian is dynamically unstable, i.e., it is an inverted harmonic oscillator in some direction. One may check that this is equivalent to $J > 1$, i.e., to the fixed point $(1,0,0)$ being a saddle in the LMG model. To do this, a useful trick is to consider the rotation 
\begin{align}\label{eq:rotation}
   &  q_C = \sqrt{x} q_A + \sqrt{1-x} q_B  \\
    &  q_D =\sqrt{1-x} q_A - \sqrt{x} q_B
\end{align}  
and similarly for $p$. Then we find that the Hamiltonian acts on $C$ and $D$ independently, as follows:
\begin{equation}
    H = - H_{0D} - H_C \,,\, H_C =  H_{0C} - \frac{J}2 q_C^2  \,. \label{eq:HCD}
\end{equation}
Namely, the $D$ subsystem is always a trivial oscillator, and the $C$-subsystem becomes an inverted oscillator iff $J > 1$. A similar rotation $A, B\to C,D$ will be used in the DHS examples below.

It is worth remarking the structure of the effective Hamiltonian \eqref{eq:Heff-TSS} and \eqref{eq:H0-TSS}: it is a sum of $H_0$, with two copies of the same Hamiltonian acting on the two subsystems, and a term coupling them, $\propto (\sqrt{x} O_A + \sqrt{1-x} O_B)$, where $O_A$ is an operator whose Green functions under the time evolution of $H_{0,A}$ match $G$ of section~\ref{sec:onloop}. 

\subsubsection{DHS examples}
Having illustrated the method in a wellstudied example, we come to examples 2 and 3 of section~\ref{sec:onloop}, which are in the DHS. In fact, their reverse-engineering problem admits a similar solution to the previous section. The main difference is that we need two degrees of freedom per subsystem:
\begin{align}
  &  H = H_{0A} + H_{0B}  + \frac{J}{2} (\sqrt{x} O_A + \sqrt{1-x} O_B )^2 \,, \label{eq:effH-DHS} \\
 &  H_{0} =  \frac12 (q_1^2 + p_1^2 - q_2^2 - p_2^2) \,,\, O = \sqrt{\kappa} (q_1 + q_2) \,,
\end{align}
where $\kappa$ will be determined later {and where the $A$ or $B$ index is implicit in the second line}. The initial state will be the ground state of 
$$ q_{1A}^2 + p_{1A}^2 + q_{2A}^2 + p_{2A}^2 +  (A\to B) \,. $$
Then, following the same steps as in the previous section, we are brought to calculate the Green function of $O$ (on the Keldysh contour): 
\begin{equation}
    \left<  0 \vert \mathcal{T}_{\sigma_1\sigma_2} O(t_1) O(t_2) \vert 0 \right> = \kappa \cos(t_{12}) \,,
\end{equation}
independently of $\sigma_1, \sigma_2$. This must match \eqref{eq:Gcase2} and \eqref{eq:Gcase3} for example 2 and 3 respectively, fix the value of $\kappa$
\begin{equation}
    \kappa = \begin{cases}
         1/3  &  \text{Example 2} \\
         1/4 &  \text{Example 3} \,. \\
    \end{cases}
\end{equation}

Now we can analyze the dynamical stability of $H$ using the same rotation method as above [see \eqref{eq:rotation}-\eqref{eq:HCD}]. The nontrivial $C$-subsystem Hamiltonian is 
\begin{align}
    H_C &= \frac12 (q_1^2 + p_1^2 - q_2^2 - p_2^2) +  \frac{\kappa J}{2} (q_1 + q_2)^2 \,.
\end{align}
where we dropped the subscript $C$ on the right hand side. One can check that, for any $\kappa J$, the dynamical matrix as in 
\begin{equation}\label{eq:dynamics-mat}
\frac{\mathrm{d}}{\mathrm{d} t}\begin{pmatrix} 
    q_1 \\ p_1 \\ q_2 \\ p_2     
    \end{pmatrix} 
    =   \begin{pmatrix} 
  0  & 1 & 0 & 0\\
    -1 + \kappa J & 0 & \kappa J & 0 \\
   0 & 0 &  0 & -1 \\
    \kappa J & 0 & 1 + \kappa J & 0
    \end{pmatrix}   \begin{pmatrix} 
    q_1 \\ p_1 \\ q_2 \\ p_2     
    \end{pmatrix} 
\end{equation}
has two Jordan blocks of size $2$ with eigenvalues $\pm i$. So there are no Lyapunov exponents with positive real part. This implies the absence of exponential growth. Yet, the Jordan blocks implies that the phase space distribution (associated with the bosonic Gaussian state) is elongated linearly in $t$ in two independent directions. Therefore we expect a logarithmic growth of entanglement entropy (see \cite{bianchi18,lerose-pappalardi2020pra} for detailed explanation):
\begin{equation} S_n \sim 2 \ln t  \label{eq:loggrowth-LMG}
\end{equation} 
for both DHS examples.  Note that this holds for any nonzero value of $J$, regardless of the existence or not of a saddle point in the TSS phase space. This is in contrast with the TSS example, where the saddle point results in a linear entanglement growth. 

It is amusing to remark that our approach reduced the difference between DHS and TSS initial product states to an innocent looking modification of the effective Hamiltonian [compare \eqref{eq:Heff-TSS} and \eqref{eq:effH-DHS}], which however leads to a qualitative change in the entanglement growth behavior. Also, the TSS effective Hamiltonian has a physical interpretation: it describes the linearized dynamics around the semiclassical trajectory. However, we cannot find an analogous interpretation for  the DHS effective Hamiltonian (to begin with, there is no semiclassical trajectory).  


\subsection{The Euler top}\label{sec:euler-ent}
So far we have focused on the LMG model for concreteness. However, the approach we used applies to any Hamiltonian at most quadratic in the collective spin variables $\mathcal{S}^a$. To illustrate, we shall briefly sketch how to apply the approach to the Euler top, $H = S \sum_a J_a \mathcal{S}_a^2 / 2$, where $J_a > 0$ for all $a$. Instead of repeating every step in detail, we shall highlight the main differences with LMG. 

To start, the path integral will involve a field with three components $\varphi \to \varphi_a$, $a = x, y,z$ (in addition to the replica and forward/backward indices), in order to decouple the $J_a \mathcal{S}_a^2 / 2$ in the Euler top Hamiltonian. Now, the Euler top Hamiltonian has no term linear in $\mathcal{S}_a$. Therefore, the decoupled one-site Hamiltonian $H_{\varphi} = \sum_a \varphi_a S_a$ \textit{vanishes} when evaluated at $\varphi_a^{\text{cl}} = 0$ (which one checks is still a classical saddle point). As a consequence, the Green functions that appear in the determinant are particularly simple:
\begin{align}
& 2  \overline{\left< \mathcal{T}_{\sigma_1\sigma_2} S^a(t_1)S^b(t_2)  \right>_s -  \left< S^a(t_1) \right>_s  \left< S^b(t_2) \right>_s}^s  \nonumber \\ = &
 \frac12 (\delta_{ab} - 
\overline{ s_a s_b }^s) =: g_{ab} \label{eq:gab}
\end{align}
for any $\sigma_1$, $\sigma_2$, $t_1$, $t_2$. Therefore, the effective Hamiltonian will have the form similar to \eqref{eq:effH-DHS}. It will act on three boson modes per half system, with positions and momenta $q_{A,B}^a, p_{A, B}^a$, $a = x,y,z$. The initial state is the ground state of $\sum_a (q_A^a)^2 + (p_A^a)^2 + (A\to B)$. To recreate the covariance matrix $g_{ab}$, we shall find $\{\tilde{p}_a\}$ which are linear combinations of $\{p_a\}$ such that  
\begin{equation}
    \left< 0 \vert \tilde{p}_a \tilde{p}_b \vert 0\right> = g_{ab}
\end{equation} 
Then the effective Hamiltonian is given by
\begin{equation}
    H = \sum_a \frac{J_a}{2} (\sqrt{x} \tilde{p}_A^a + \sqrt{1-x} \tilde{p}_B^a)^2 \,.
\end{equation}
 It follows then that the dynamical matrix [analogue of \eqref{eq:dynamics-mat}] will have 3 Jordan blocks of size $2$ eigenvalue $1$ (representing $\partial_t \tilde{q}_a \propto \tilde{p}_a$, $\partial_t \tilde{p}_a = 0,$, where $\tilde{q}_a$ are the canonical conjugate of $\tilde{p}_a$). As a consequence, we have
\begin{equation}
     S_n \sim m \ln t  \label{eq:entlog-Euler}
\end{equation}
where $m$ is the number of nonzero $J_a$'s. Note that it is not important whether the $J_a$'s are distinct; in particular, even if $J_a$ are all equal (and nonzero), we still have $S_n \sim 3\ln t$, although the dynamics is trivial in every sector of fixed total spin. This is in stark contrast with the OTOC, which depends qualitatively on (and only on) the difference between $J_a$'s. In particular, when $J_x = J_y = J_z \ne 0$, the entanglement entropy grows but no OTOC does since $\mathbf{O}(t) = \mathbf{O}$ has trivial time evolution. 

We observe that in all DHS examples considered so far, the entanglement growth is (at most) logarithmic in time. We surmise that the logarithmic growth of entanglement is generic in uniform all-to-all models starting from a DHS product state. This conjecture will be in more detail studied elsewhere.

\subsection{Finite $N$ numerics}\label{sec:finiteN}
The semiclassical theory developed so far is exact in the limit of $N\to\infty$ with $t$ fixed. However, for finite $N$, the (Renyi or von Neumann) entanglement entropy is bounded, and its growth must thus saturate at a time parametrically long in $N$. In this section, we study numerically this saturation process.

For a quench from a TSS initial state, the entanglement entropy is bounded by the log of the TSS dimension: $ S_n \lesssim \ln N$. Numerical study has observed that this bound is often asymptotically saturated~\cite{richter,lerose-pappalardi2020pra}. The entanglement entropy growth thus follows the semiclassical prediction, until saturating at a value $\propto \ln N$. 

By contrast, for a quench in the DHS, the entanglement entropy is \textit{a priori} only bounded by the {log of} the Hilbert space dimension of the subsystem: $S_n \lesssim N$, and we would naively expect this bound to be asymptotically saturated. However, our numerical results indicate the contrary.

\begin{figure}
    \centering
    \includegraphics[width=\columnwidth]{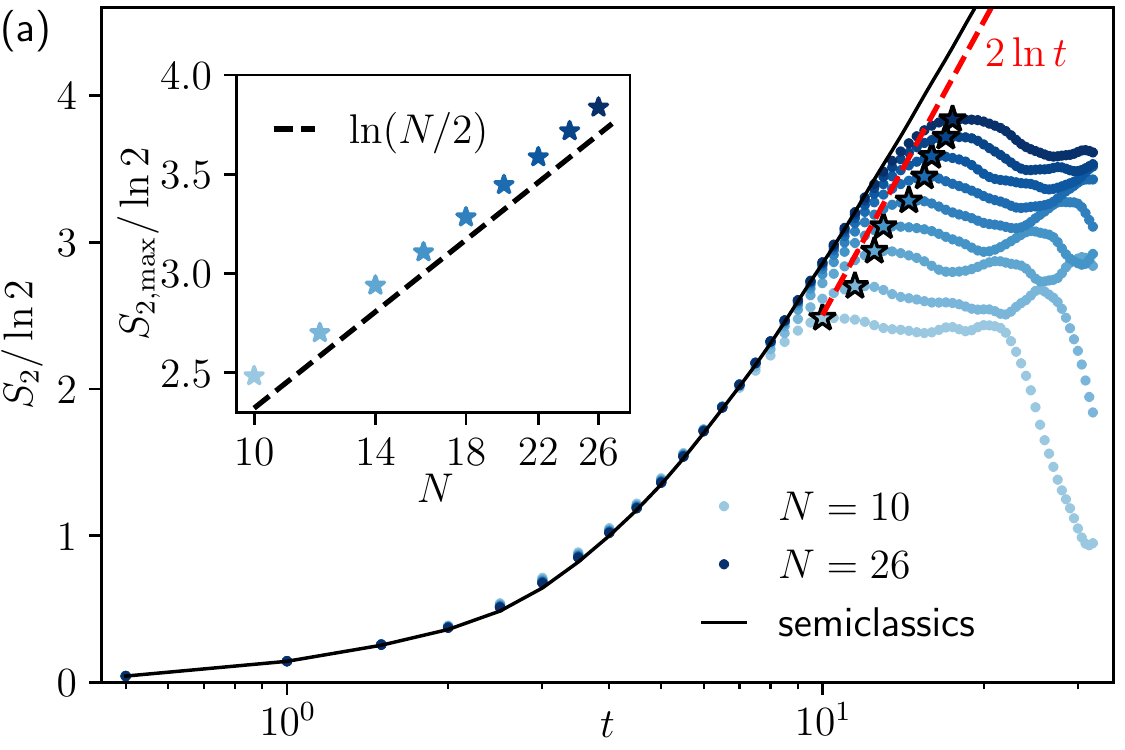}
     \includegraphics[width=\columnwidth]{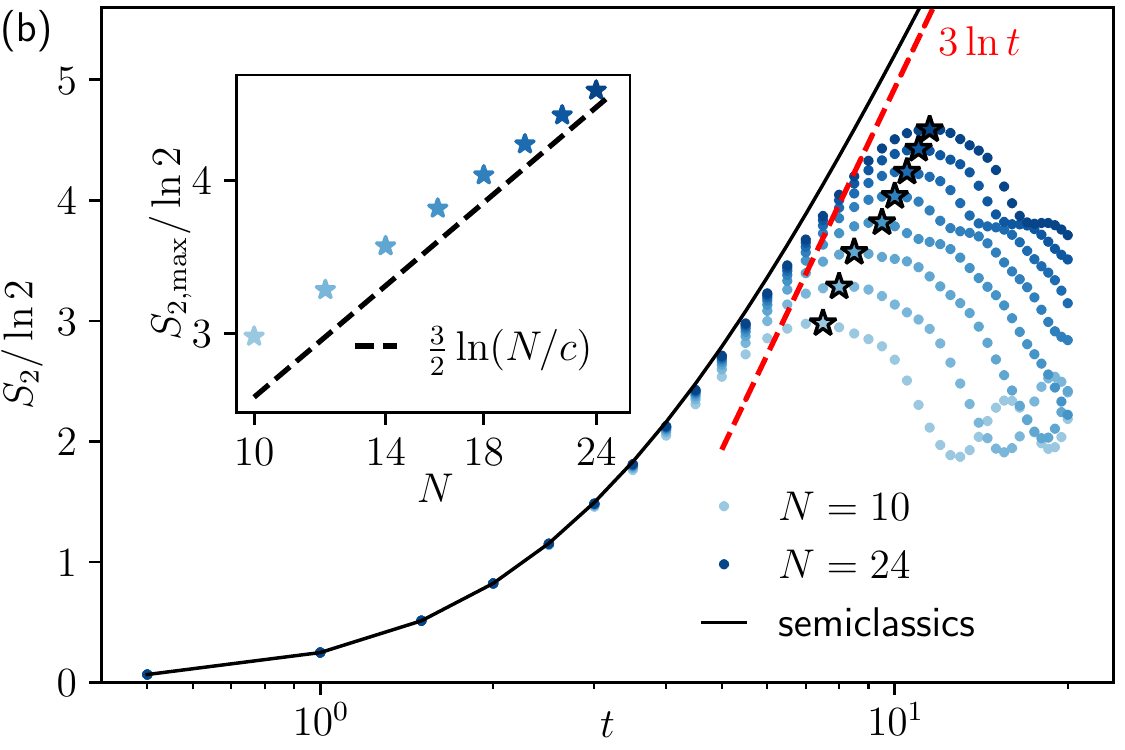}
    \caption{
    Entanglement entropy (Renyi, $n=2$, equal bi-partition) in a quantum quench from a DHS product state~\eqref{eq:defPsi0}, with $s_1 = s_{N/2+1}, s_2 = s_{N/2+2}, \dots$ equally spaced on the great circle $\{x=0\}$ of the Bloch sphere (example 3 in Section \ref{sec:onloop}). The two panels correspond to two Hamiltonians. \textbf{(a)} LMG model $H = \sum_j S_j^x + J \sum_{jk} S_j^z S_k^z / (2S) $, where $S = N/2$ and $J=2$.  Note that we need to use the ``$1/N$ normalization''~\eqref{eq:HandSa}. 
    \textbf{(b)} The isotropic Euler top $H=\sum_{jk,a} S_j^a S_k^a / (2S)$.  \textbf{Main}. the finite $N$ numerical data (obtained from direct simulation) are compared to semiclassical prediction (obtained from the one-loop determinant \eqref{eq:Renyidet} discretized in time), as well as to the asymptotic logarithmic growth prediction $S_2 \sim c\ln t$~\eqref{eq:loggrowth-LMG}, \eqref{eq:entlog-Euler}, respectively. For each $N$, we measure the saturation value $S_{2,\max}$ as the first local maximum of $S_{2}$ (indicated by a star). \textbf{Inset}. The $N$ dependence of $S_{2,\max} $, compared to $\frac{c}2\ln N$~\eqref{eq:S-saturation}. The simulated time evolution is Trotterized (kicked) with $\delta t = 0.5$, see \eqref{eq:floquet} and Appendix~\ref{app:numerics1}. The semiclassical prediction has the same time discretization, and should thus match exactly the Trotterized numerics {in the large $N$ limit}.  }
    \label{fig:entanglement}
\end{figure}
We simulated directly (see Appendix~\ref{app:numerics1} for methods) two of the quantum quenches studied above, and computed the $n=2$ Renyi entanglement entropy in systems with $N \le 26$. For the sake of numerical efficiency, we simulated trotterized (kicked) variants of the Hamiltonians, see \eqref{eq:floquet} below. The results are shown in Figure~\ref{fig:entanglement}. As a check, we compared them to the semiclassical prediction, calculated as the one-loop determinant~\eqref{eq:Renyidet} discretized in time. As expected, for fixed $t$, the numerical result converges to the semiclassical prediction as $N$ increases; the $c \ln t$ growths predicted in Section~\ref{sec:effH} are also observed, with the correct pre-factor. We then turn our interest to the saturation regime, and extracted the saturation value $S_{\max}$ as the first (in time) local maximum of the the entanglement entropy. We observed in both cases that 
\begin{equation} \label{eq:S-saturation}
    S_{\max} \sim \frac{c}2 \ln N \,,\, S \sim  c \ln t \,.
\end{equation}
As a result, the entanglement saturation time scale is
\begin{equation}
    t_{\text{ent}} \sim N^{\frac12} \,.
\end{equation}
We simulated the dynamics for longer times and did not see any further growth beyond $t_{\text{ent}}$ (in the parlance of Section~\ref{sec:finiteNOTOC}, the entanglement growth follows a ``normal scenario''). 

The precocious saturation of entanglement growth from a DHS product state is intriguing, and worth a few remarks. The saturation does not rely on the absence of chaos in the (TSS) phase space dynamics. Indeed, we simulated kicked deformations of the above models; for LMG for example, the time evolution by $\delta t$ can be described by a Floquet unitary
\begin{equation}\label{eq:floquet}
    U = e^{-i \sum_j S_j^x \delta t } e^{-i \frac{1}{2S} \sum_{jk}S_j^z S_k^z  \delta t } \,. 
\end{equation}
As one increases $\delta t$, the TSS phase space dynamics becomes (partially) chaotic, yet \eqref{eq:S-saturation} still holds. In fact, the field theory method above can be readily adapted to treat the kicked Hamiltonians. It suffices to replace continuous time integrals by discrete sums (which is done in the numerics). The effective free boson Hamiltonian also becomes the kicked variant (constructed similarly as that of the original Hamiltonian), and one can check that the dynamical matrix has an identical Jordan block structure (with stable eigenvalues) in all cases studied above. On the other hand, the uniform all-to-all interaction seems crucial to maintaining the low-entanglement. For example, we observed that adding spatial disorder or space-time noise (of any nonzero amplitude) to the $\sum_j S_j^x$ term in \eqref{eq:floquet} makes the entanglement grow eventually to a volume law. 

 The main lesson of this section is that the entanglement dynamics from a DHS product state has nothing to do with that of the OTOCs and autocorrelation functions: they are governed by different time scales, and have distinct large $N$ descriptions. This is less surprising than it sounds. The OTOC involves time-evolved few-body operators. Meanwhile, the Renyi entanglement entropy is equal to the expectation value of the partial swap operator, which acts on $O(N)$ sites at once. The dynamics of such an operator is not described by the large $N$ phase space method which we used to calculate OTOC and auto-correlation function, and which is only valid when the operator size is much smaller than $N$. Hence, there is generically no reason to expect a close relation between entanglement growth and OTOC (of few-body operators). In this regard, the situation in the TSS is rather exceptional. There, it is possible to identify a Lyapunov spectrum, which governs both OTOC and entanglement entropy. If such a theory were to exist in the DHS, it must be of a distinct form.

\section{Conclusion}
The Hilbert space of spin systems with uniform all-to-all interaction is fragmented into sectors of various conserved total spin. The totally symmetric space (TSS) has maximal total spin $S = N/2$ and a well studied semiclassical $N\to\infty$ limit. Here, we unveiled remarkable properties of the all-to-all quantum dynamics in the deep Hilbert space, characterized by $S = O(\sqrt{N})$. The growth of local operators (as measured by OTOC and K-complexity) and that of entanglement from a product state have parametrically separated time scales. The initial stage of both can be described by large $N$ theories, each of which modifies the TSS semiclassical theory in a distinct way. In paradigmatic examples, the OTOC has a super-exponential initial growth followed by a slow saturation, so the uniform all-to-all models in the DHS are \textit{not} fast scramblers. The entanglement dynamics generically exhibits a logarithmic in time growth (even in the presence of classical chaos), and saturates at a value which is logarithmic in system size, without reaching a volume law. These results are summarized in Table~\ref{tab:summary} (with the example of the Euler top model), and compared to the TSS analogues.  

These findings paint the DHS as a rather exotic world, much different from the TSS. How can this be possible, considering that all the sectors of the Hilbert space are just like the TSS except with different total spin? The crux is that the decomposition into spin sectors has a complex relation with spatial locality (or the tensor product structure of the Hilbert space), except for the TSS. Therefore, physical observables away from the TSS typically involve several sectors, and it is not obvious to isolate their respective contributions. For instance, the entanglement growth from a DHS product state cannot be symmetry resolved~\cite{Laflorencie_2014,moshe18,Bonsignori_2019,symrespage}.

This being said, the early saturation of the entanglement growth suggests that the DHS is sill fragmented in some more intricate way. In particular, a DHS product state and a Haar-random state belong to different disconnected fragments of the DHS: no all-to-all dynamics can transform one to the other. In fact, the DHS product states are ``half-deep'': they have a DHS OTOC behavior, but resemble rather the TSS ones in terms of entanglement. On the other extreme, there are a large set of spin singlet ($S = 0$) states that do not evolve at all, and thus are disconnected from everything else. Are there other fragments? Characterizing the inner structure of the DHS is an interesting problem. 

How generic is the deep Hilbert space physics? 
Although we have focused on uniform all-to-all models in this work, we expect most of our results to apply to systems with sufficiently long-range interactions~\cite{pappalardilerosePRR,pappalardi18} or sufficiently weak randomness~\cite{monika-ehud,Monika}.
(A plausible exception will be the long-time entanglement saturation at a sub-volume law, see above.)
We remark also that models that resemble SYK could have deep Hilbert space phenomena as well. In fact, it may be possible to interpolate between models with a DHS and those with maximal chaos and a holographic dual, for example by considering the ``low-rank'' SYK model~\cite{lowrank}. 

Thus, there seems to be a broad class of ``weakly chaotic'' long-range interacting systems hosting a deep Hilbert space, where the co-existence of disparate time scales can dramatically affect the many-body quantum dynamics. We hope to further explore these deep Hilbert spaces in future work. 

\begin{acknowledgments}
T. S. acknowledges the support of the
Natural Sciences and Engineering Research Council of
Canada (NSERC), in particular the Discovery Grant
(No. RGPIN-2020-05842), the Accelerator Supplement
(No. RGPAS-2020-00060) and the Discovery Launch
Supplement (No. DGECR-2020-00222). 
X.C. was supported by CNRS and ENS. 
\end{acknowledgments}



\bibliography{main}

\appendix

\section{Reminder on the recursion method and K-complexity}\label{app:lanczos}
For the reader's convenience, we recall some facts of the recursion method used in Section~\ref{sec:k-review} and \ref{sec:k-dhs} on K-complexity. See also~\cite{viswanath2008recursion,hyp}.

We first recall the Lanczos algorithm. Its input is the Liouvillian, $\mathbb{L} = [H, \cdots]$, the inner product of the operator space, e.g., $(A | B) = \Tr[\rho_{\infty}A^\dagger B]$~\eqref{eq:infiniteTinner}, and an initial operator $O_0$. We assume that $O$ is Hermitian and normalized $(O_0 \vert O_0) = 1$. The output is a sequence of Lanczos coefficients $b_n$.

One starts by computing $A_1 := \mathbb{L} O_0 $ and $b_1 := \sqrt{( A_1 | A_1 )}$. We normalize $A_1$ and define $O_1 := b_1^{-1} A_1$. The rest of the Lanczos coefficients and Krylov basis elements can be iteratively defined as:
\begin{align}
    A_n &= \mathbb{L} O_{n-1} - b_{n-1} O_{n-2} \nonumber \\
    b_n &= \sqrt{( A_n | A_n)} \nonumber \\
    O_n &= b_n^{-1} A_n \,,\, n \ge 2 \,.
\end{align}
The algorithm halts when $b_n = 0$ for some $n$. The resulting Krylov basis $\{ O_n \}$ is orthonormal: $( O_n | O_m ) = \delta_{nm}$. 
Furthermore, the Liouvillian super-operator is tri-diagonal in this basis, with the diagonal entries being zero and the subdiagonal entries the Lanczos coefficients: $( O_n | \mathbb{L} | O_m ) = b_n \delta_{n, m-1} + b_m \delta_{n, m+1}$.

The asymptotic behavior of the Lanczos coefficients are related to the high-frequency tail of the  spectral density $\rho(\omega)$ through its moments defined as
\begin{equation}
    \mu_{2n} = \int \omega^{2n} \rho(\omega) \frac{\mathrm{d} \omega}{2\pi} \,.
\end{equation}
A saddle point approximation of the $\omega$ integral then shows that  
\begin{equation}
    \rho(\omega) \sim \exp(- c |\omega|^{a} ) \implies \mu_{2n} \sim n^{\frac{2n}a} \exp(c' n) \,. \label{eq:rho-mu}
\end{equation}
Meanwhile,  it is also known that~\cite{viswanath2008recursion,avdoshkin}, if the Lanczos coefficients have a power-law asymptotics,
\begin{equation}\label{eq:mub2n}
   \mu_{2n} \sim b_1^2 \dots b_n^2 e^{Cn} \,.
\end{equation}
Combining \eqref{eq:rho-mu} and \eqref{eq:mub2n}, we obtain a dictionary 
\begin{equation}
    \rho(\omega) \sim \exp(- c |\omega|^{a} )  \Leftrightarrow b_n \sim n^{\frac1a} \,.
\end{equation}
The equivalence is not rigorous and depends on technical conditions on $b_n$~\cite{avdoshkin}. But it is still useful to obtain predictions on the growth rate of $b_n$ from the spectral density. In the models considered here, the predictions are checked to be correct, see  for example Figure~\ref{fig:dhs-lanczos}.

\section{K-complexity of the LMG model in the TSS}\label{app:LMG}

The K-complexity growth rate $\omega_0$ for the LMG model $h = x + J z^2 / 2$ in the TSS has been explicitly computed in Ref.~\cite{Bhattacharjee_2022}. In fact, that work considered the micro-canonical ensemble with an energy $E \in (-1, (J + 1/J)/2)$, and showed that 
\begin{equation} \label{eq:omega0_LMG}
    \omega_0(J,E) =  \frac { \sqrt{\sqrt{J^2- 2E J+1}+ E J-1}}{ 2 \sqrt{2} \,\mathsf{K}\left(\frac{1- E J +\sqrt{J^2- 2E J+1}}{1- E J -\sqrt{ J^2-2 E J+1}}\right)}
\end{equation}
where $ \mathsf{K}$ is the complete elliptic integral of first kind [see Eq. (4.5) thereof; note that the model in that work is parametrized a bit differently as $x + J z^2$] The TSS K-complexity growth rate is obtained by maximizing with regard to $E$. To find the asymptotics at large $J$, let us write $ E = k J$, so that $k \in (-1, 1/2 + O(1/J))$. Then the numerator of \eqref{eq:omega0_LMG} $\sim \sqrt{k} J$. For the denominator, the argument of $\mathsf{K}$ tends to $1$ as $1 - c / J + O(1/J^2)$ with  $c=2 \sqrt{1-2k}/k$, and $\mathsf{K}(1 + x) \sim - \frac12 \ln x$ is a log singularity. Therefore we have 
\begin{equation}
     \omega_0(J, k J) \sim \frac{J}{ \ln J} \,.
\end{equation}
Hence, K-complexity growth rate in the TSS behaves as $\omega_0 \sim J / \ln J$ at large $J$. In the DHS context of the main text, $J = r$, so $\omega_0(r) \sim r / \ln r$. 

\section{Phase space representation of $S_i^a(t)$}\label{app:singlespin}

In the main text we focused on the time evolution of a DHS collective spin variable under an all-to-all Hamiltonian with $1/\sqrt{N}$ normalization. In the large $N$ limit, this can be described semiclassically, in terms of a function on the phase space $\mathbb{R}^3$.  Here, we discuss the time evolution of a single-site operator, $S_i^a(t)$, under the same Hamiltonian. Such an operator appears in the ``standard'' definition of the OTOC, see~\eqref{eq:standardOTOC}. Our goal here is to show \eqref{eq:Siat} that is used to argue that the standard OTOC is no much different from the one we focus on. 

The time evolution of $ S_i^a(t)$ is generated by the application of the ``Liouvillian'' $i[H, \cdot]$ where $H = \sqrt{N} h (\mathbf{S}_a)$. Now, observe that the single-spin operators behave similarly as the collective ones with respect to the commutation with collective spins:
$$ \sqrt{N}[\mathbf{S}_a, S_i^b] = i \epsilon_{abc} S_i^c\,, \sqrt{N}[ \mathbf{S}_a, \mathbf{S}_b] = i \epsilon_{abc} \mathbf{S}_c \,. $$
Therefore, we can treat $S_i^a$ as classical variables that evolve under the Poisson bracket just like $\mathbf{S}_a$. Therefore,  the time evolution $S_i^a(t)$ will be just like that of $\mathbf{S}_a(t)$, except that if we view the latter as a power series in the collective spins, exactly one factor in each term is ``marked'', i.e., replaced by the corresponding single site operator $\mathbf{S}_b\to S_i^b$. Now, it is not hard to see that the operation of ``marking'' a factor can be achieved by the differential operator $\sum_b  S_i^b \partial_{\mathbf{S}_b}$. Hence, we have 
$$  
S_i^a(t) = \sum_b S_i^b \partial_{\mathbf{S}_b} \mathbf{S}_a(t) \,,
$$
which is \eqref{eq:Siat} in the main text. One may think of the phase space representation of $S_i^a(t)$ as a vector field on the phase space, with components $\partial_{\mathbf{S}_b} \mathbf{S}_a(t)$, $b = x,y,z$. In this sense it is more involved object than a time evolved collective spin, which is a scalar function on the phase space.

\section{Operator numerics in the DHS}\label{app:numerics}
In this Appendix we detail the numerical method used to calculate Lanzcos coefficients and OTOCs in the DHS, first in finite $N$ systems and then in the $N\to\infty$ limit. We will use the latter to make connection with Hermite polynomials in the next Appendix. 

We first consider a system of finite $N$. The main idea is to represent the evolving operator $|\mathbf{O}(t))$ as a vector in the space of symmetric operators, using the computational basis of symmetrised Pauli strings~\eqref{eq:ellmndef}: 
\begin{align} \label{eq:lmnapp}
    &| \ell, m, n )  \\ := &
    \binom{N}{\ell,m,n}^{-\frac12} \sum^*_{\substack{i_1 < \dots < i_\ell \\ 
    j_1 < \dots < j_m \\ k_1 < \dots k_n
    }} |\sigma_{i_1}^x  \dots _{i_\ell}^x \sigma_{j_1}^y  \dots \sigma_{i_m}^y  
     \sigma_{k_1}^z  \dots \sigma_{k_n}^z ) \,,\nonumber  \, \\
     &  \ell + m + n \le N
\end{align}
where the sum $\sum^*$ is over indices such that $\{i_1, \dots, i_\ell\}$, $\{j_1, \dots, j_m\} $, $\{k_1, \dots, k_n\} $ are mutually disjoint. The dimension of the operator space is $(N+1)(N+2)(N+3)/6$. The computational basis is orthonormal under the infinite-temperature inner product $(A | B) = \mathrm{Tr}[\rho_{\infty} A^\dagger B]$, and its elements have definite operator size. Therefore, computing the inner product and the OTOC are straightforward tasks. 

It remains to see how to apply the Liouvillian $\mathbb{L} := [H, \dots]$, which is needed to implement the Lanczos algorithm and calculate the time evolution. Since $H$ is an all-to-all Hamiltonian in the form of \eqref{eq:HDHS} (with the $1/\sqrt{N}$ normalization), it suffices to implement the commutator and the anticommutator with respect to $\mathbf{S}_a = N^{-1/2} \sum_j S_j^a$:
\begin{equation}\label{eq:LandM}
  \mathbb{L}_a O := \sqrt{N} [\mathbf{S}_a,  O ]  \,,\,  \mathbb{M}_a O :=  \{ \mathbf{S}_a,  O \} \,.
\end{equation}
(Here $\{A, B\} = A B + B A$ denotes the anticommutator, not to be confused with the Poisson bracket $\{\cdot , \cdot\}_{\text{P.B.}}$.) Indeed, the commutator with respect to a two-body term can be obtained by composition, for example:
\begin{equation}
  [ \sqrt{N}  \mathbf{S}_a^2, O] = \mathbb{M}_a  \mathbb{L}_a  O \,.
\end{equation}
Note that $ \sqrt{N}  \mathbf{S}_a^2 $ follows the $1/\sqrt{N}$ normalization of the Hamiltonian \eqref{eq:HDHS}.

We now describe $ \mathbb{L}_a $ and $ \mathbb{M}_a $ in the computation basis. For definiteness, we focus on $a = x$ (the other cases follow from cyclic permutation). We claim that:
\begin{align}
     \mathbb{L}_x  | \ell, m,n ) = 
    &  i \sqrt{m (n+1)} \, | \ell, m-1, n+1 ) \nonumber \\ 
     - & i \sqrt{m (n+1)} \, | \ell, m+1, n-1 ) \label{eq:Lx} \\
     \mathbb{M}_x | \ell, m,n ) = &  \sqrt{\frac{h (\ell + 1)}N} \, | \ell + 1, m, n ) \nonumber \\
     + & \sqrt{\frac{(h + 1) \ell}N} \, | \ell - 1, m, n ) \label{eq:Mx}
\end{align}
where 
\begin{equation}
    h = N - \ell - m - n 
\end{equation}
is the number of sites occupied by identity. The above results can be checked by an explicit calculation following the definitions of $\mathbf{S}_a$ \eqref{eq:SaDHS}, and of $|\ell,m,n)$ \eqref{eq:lmnapp}. We may also understand them (in particular the square roots) by viewing $|\ell, m, n)$ as a Fock state in a fictitious bosonic system of four flavors $x, y, z, I$, which correspond to the three Pauli's and the identity and are occupied by $\ell, m, n, h$ bosons, respectively. The Bose statistics come from the symmetrization in \eqref{eq:lmnapp}. In this regard, $ \mathbb{L}_x$ acts as a hopping operator between $y$ and $z$ flavors, and $\sqrt{N} \mathbb{M}_x$ between $x$ and $I$. Both follow directly from Pauli algebra. The square roots in \eqref{eq:Lx} and \eqref{eq:Mx} arise naturally from boson algebra. 

Now, we can readily take the $N \to \infty$ limit. Indeed, observe that all the matrix elements of \eqref{eq:Lx} and \eqref{eq:Mx} have a $N\to\infty$ limit (with $\ell, m,n$ fixed): those of $ \mathbb{L}_x $ does not depend on $N$, and 
\begin{equation}\label{eq:MxNinf}
     \mathbb{M}_x | \ell, m,n )  \stackrel{N\to\infty} =  \sqrt{{\ell + 1}} \, | \ell + 1, m, n ) +
 \sqrt{{ \ell}} \, | \ell - 1, m, n ) \,.
\end{equation}
Therefore, to calculate the operator dynamics in the large $N$ limit amounts essentially to replacing \eqref{eq:Mx} by \eqref{eq:MxNinf}. Another change is that the space of symmetric operator becomes infinite-dimensional, spanned by all $| \ell, m, n)$, $\ell, m , n \ge 0$. In practice, we apply a numerical truncation, $\ell + m + n \le M$ when implementing the Lanczos algorithm, so that the first $M-1$ Lanczos coefficients are exact (assuming that the Hamiltonian involves at most two-body interaction and the initial operator is one-body). The OTOC numerics is always implemented at finite $N$. 



\section{Relating to Hermite polynomials}\label{app:hermite}
In this appendix we establish the connection between the symmetric operators in the DHS and Hermite polynomials. For this, we first gather some relevant facts about the Hermite polynomials, or equivalently, basic quantum mechanics of the harmonic oscillator. We define the Hermite polynomials $\chi_n(x)$ as polynomials of degree $n$ satisfying the orthonormality relation 
\begin{equation} \label{eq:hermiteOrthonormal-app}
\int \chi_m(x) \chi_n(x) e^{-2x^2} \frac{ \mathrm{d} x}{\sqrt{\pi/2}} = \delta_{mn} \,.
\end{equation}
These conditions uniquely define $\chi_n$, which can be obtained by a Gram-Schmidt process (or Lanczos algorithm). In other words, 
\begin{equation}
    \psi_n(x) := \chi_n(x) e^{-x^2} 
\end{equation} 
is the wave-function of the $n$-th eigenstate of a quantum harmonic oscillator with Hamiltonian 
\begin{align}
    H =   \frac12  (-\partial_x^2 + 4 x^2) = 2 \left(a^\dagger a +  \frac12 \right) \,,
\end{align}
such that 
\begin{equation}
     H \psi_n(x) = (2 n + 1) \psi_n(x) \,.
\end{equation}
[We note in passing that the last equation is equivalent to the differential equation \eqref{eq:diffeq-Hermite} satisfied by the Hermite polynomials.] The ladder operators are given by $ a = x +  \partial_x / 2$ and $a^\dagger = x -  \partial_x / 2  $, so that $2 x = a + a^\dagger$. The usual boson algebra then implies the three-term recurrence of the Hermite polynomials:
\begin{equation}
    2 x \chi_n(x) = \sqrt{n + 1} \, \chi_{n+1}(x) + \sqrt{n} \, \chi_{n-1}(x) \,.
\end{equation}
This equation is to be compared with \eqref{eq:MxNinf}: their similarity allows us to relate the symmetric operators $|\ell, m, n)$ to Hermite polynomials. Indeed, consider the action of the anticommutator $\mathbb{M}_a$~\eqref{eq:LandM} on $  \chi_\ell(\mathbf{S}_x)  \chi_m(\mathbf{S}_y) \chi_n(\mathbf{S}_z) $ in the large $N$ limit. There, we can consider the operators $ \mathbf{S}_z$ to commute with each other [their commutator is $O(1/\sqrt{N})$]. So the action of $\mathbb{M}_a$ is simply multiplying by $2 \mathbf{S}_a$, and we have 
\begin{align}
   & \mathbb{M}_x  \left[  \chi_\ell(\mathbf{S}_x)  \chi_m(\mathbf{S}_y) \chi_n(\mathbf{S}_y)  \right] \nonumber  \\ 
    =& 2 \mathbf{S}_x  \chi_\ell(\mathbf{S}_x)  \chi_m(\mathbf{S}_y) \chi_n(\mathbf{S}_y)   \nonumber \\
    =& ( \sqrt{\ell+1} \, \chi_{\ell+1}(\mathbf{S}_x) + \sqrt{\ell}  \, \chi_{\ell-1}(\mathbf{S}_x))  \nonumber \\
     & \times \chi_m(\mathbf{S}_y) \chi_n(\mathbf{S}_y)  \nonumber
\end{align}
and similarly for $\mathbb{M}_y$ and $\mathbb{M}_z$. Comparing to \eqref{eq:MxNinf}, we conclude that $ | \ell, m, n ) $ satisfy the same recurrent relations as $  \chi_\ell(\mathbf{S}_x)  \chi_m(\mathbf{S}_y) \chi_n(\mathbf{S}_z)$. Moreover, since $\chi_0(x) =1$, the ``initial conditions'' are the same $ |0,0,0) =  \chi_0(\mathbf{S}_x)  \chi_0(\mathbf{S}_y) \chi_0(\mathbf{S}_z) $. So one can readily show by induction that the two sequences of operators must be identical:
\begin{equation}
    | \ell, m, n )  =  | \chi_\ell(\mathbf{S}_x)  \chi_m(\mathbf{S}_y) \chi_n(\mathbf{S}_z) ) \,,
\end{equation}
as we claimed in \eqref{eq:ellmnisHermite}.

\section{Generating function of DHS collective variables}\label{app:traces}
\noindent \textbf{Derivation of \eqref{eq:Mua}}. We use the fact that both $\rho_{\infty}$ and $e^{u_a \mathbf{S}_a}$ factorize, hence
\begin{equation} \label{eq:D1}
    M(\{u_a\}) = \Tr[\rho_{\infty} \prod_a e^{u_a \mathbf{S}_a} ] 
    = \left( \left< \prod_a e^{u_a S_a / \sqrt{N}}\right> \right)^N,
\end{equation}
where $\left< [\dots]\right>$ stands for the infinite-temperature on a single site, on which $S_a$'s act. In the large $N$ limit, we should expand up to second order in $1/\sqrt{N}$. Recalling that $\left<S_a\right> = 0$ and $\left<S_a S_b\right> = \delta_{ab} / 4$, we have:
\begin{align}
   &   \left<\prod_a e^{u_a S_a / \sqrt{N}}\right> \nonumber \\ =& 1 + \frac1{\sqrt{N}} \sum_a   \left< u_a S_a \right>  + \frac1{2N}   \sum_{ab}  \left< u_a u_b S_a S_b \right> + o(1/N)  \nonumber \\
   =& 1 + \frac1{8N} \sum_{a}  u_a^2  + o(1/N) \,.  
\end{align}
Plugging this back into \eqref{eq:D1}, we obtain \eqref{eq:Mua}.

\noindent \textbf{Derivation of Proposition in Section \ref{sec:prodDHS}.} We consider the similar generating function
\begin{equation}\label{eq:Mpsi}
    M_{\Psi}(\{u_a\}) := \langle \Psi \vert \prod_a e^{u_a \mathbf{S}_a} \vert \Psi \rangle \,,
\end{equation}
with respect to the product state \eqref{eq:defPsi0} 
\begin{equation*}
    \vert \Psi \rangle = \prod_{j = 1}^N \vert s_j \rangle \,,\,  \left< s_j \vert S_j^a \vert s_j \right> = s_j^a / 2  \,.
\end{equation*}
satisfying $\sum_{j=1}^N s_j = 0$. Again, the calculation factorizes, and we can write:
\begin{equation} \label{eq:Mpsi2}
    M_{\Psi}(\{u_a\}) = 
   \exp\left( \sum_j \ln \left< \prod_a e^{u_a S_a / \sqrt{N}}  \right>_j  \right)
\end{equation}
where $\left< [\dots] \right>_j = \left< s_j \vert [\dots] \vert s_j \right> $. Expand the log up to second order in $1/\sqrt{N}$ [recall $\ln (1+x) = x - x^2/2 + O(x^3)$], we find
\begin{align}
  & \ln \left< \prod_a e^{u_a S_a / \sqrt{N}}  \right>_j \nonumber \\ =&
    \frac{1}{\sqrt{N}} u_a \left< S_a \right>_j + \frac1{2N} \sum_{ab} u_a u_b ( \left< S_a S_b \right>_j - \left< S_a \right>_j 
   \left< S_b \right>_j) \nonumber \\
   =& \frac{1}{\sqrt{N}} u_a \left< S_a \right>_j +   \frac1{8N} \sum_{ab} u_a u_b (\delta_{ab} - s_j^a s_j^b) \,.  \label{eq:D5log}
\end{align}
Plugging this back into \eqref{eq:Mpsi2} give us 
\begin{equation}\label{eq:D6}
      M_{\Psi}(\{u_a\}) = \exp\left( \frac18 \sum_{ab} u_a u_b \left(\delta_{ab} - \overline{s^a s^b}^s \right)  \right) \,,
\end{equation}
where we recall that $\overline{f(s)}^s = \lim_{N\to\infty} \frac1N \sum_{j} f(s_j)$ denotes an average over the $s_j$'s. The term proportional to $1/\sqrt{N}$ does not contribute to \eqref{eq:D6} because of the condition $ \sum_j s_j = 0 $. Eq.~\eqref{eq:D6} implies that $\mathbf{S}_a$ behave indeed as centred Gaussian variables with the covariance matrix $ (\delta_{ab} - \overline{s^a s^b}^s) / 4$ \eqref{eq:prodDHS-res}, as we claimed in the main text.  

\section{Entanglement numerical methods}\label{app:numerics1}
We outline the numerical methods used to directly simulate the state time evolution under the (kicked) LMG and Euler top models. The ``kicking'' essentially means that we trotterize the Hamiltonian, but with rather large time steps, see \eqref{eq:floquet} for example; so it is more convenient numerically. They are used to generate Figure~\ref{fig:entanglement} in Section~\ref{sec:finiteN}, where the time step is $\delta t = 0.5$. Accordingly, the semiclassical prediction is calculated with a time-discretized version of the determinant~\eqref{eq:Renyidet}, with the the same $\delta t$. One can show that this gives the exact prediction for the kicked model, so we expect the semiclassical curve to match the numerical data perfectly as $N\to\infty$ with $t$ fixed.

We represent the state $\Psi$ exactly as a vector of size $2^N$, in the computational basis, where $S_j^z$ act diagonally. Thus, the time-evolution operator 
\begin{equation}
    T_z(u) =  e^{i u ( \sum_j S_j^z)^2}
\end{equation} 
can be applied straightforwardly, for any $u$. We compute its action (a phase) on each basis vector once and store the results (as a $2^N$ vector), and a time-evolution $|\Psi\rangle \to T_z(u) |\Psi \rangle$ is implemented by a vector-vector multiplication. Next, a spin rotation
\begin{equation}
     R_a(u) =  e^{i u  \sum_j S_j^a} = \prod_{j}e^{i u  S_j^a }
\end{equation}
where $a = x, y$ is implemented by acting on each site in turn. The computation on each site is a matrix multiplication between a $2 \times 2$ matrix (the exponential of a Pauli) and a $2 \times 2^{N-1}$ matrix (a reshape of $\Psi$ with the first index representing the site being acted on). 
Finally, combining $T_z$ with suitable spin rotations, we implement $T_x$ and $T_y$. Thus, we are able to implement all the time evolution parts involved in the kicked LMG or Euler top model, in a matrix free manner. The memory cost is $O(2^N)$ and the time cost is $O(2^N N)$ for each time evolution step. The data of Figure~\ref{fig:entanglement} are obtained on a laptop.

\end{document}